\documentclass[useAMS,usenatbib]{mn2e}
\usepackage{epsfig,times,graphicx,bm,amssymb,aas_macros}

\newcommand{\mk}{}
\newcommand{\mkb}{} 
\newcommand{\rd}{{\rm d}}
\newcommand{\del}{{\bm\nabla}}
\newcommand{\dv}{\del\cdot}
\newcommand{\curl}{\del\times}
\newcommand{\beq}{\begin{equation}}
\newcommand{\eeq}{\end{equation}}
\newcommand{\beqa}{\begin{eqnarray}}
\newcommand{\eeqa}{\end{eqnarray}}
\newcommand{\bcom}{}

\title[Discontinuities in MHD]{From pulsar scintillations to coronal heating:\\discontinuities in magnetohydrodynamics}
\author[Jonathan Braithwaite]{Jonathan Braithwaite \thanks{E-mail: jonathan$@$astro.uni-bonn.de}\\Argelander Institut f\"ur Astronomie, Auf dem H\"ugel 71, 53121 Bonn, Germany}
\pagerange{\pageref{firstpage}--\pageref{lastpage}}
\begin{document}\maketitle\label{firstpage}
\begin{abstract} From pulsar scintillations we infer the presence of sheet-like structures in the ISM; it has been suggested that these are current sheets. {\mk Current sheets probably play an important role in heating the solar corona, and there is  evidence for their presence in the solar wind. Such magnetic discontinuities have been found in numerical simulations with particular boundary conditions, as well as in simulations using an incompressible equation of state. Here, I investigate their formation under more general circumstances by means of topological considerations as well as numerical simulations of the relaxation of an arbitrary smoothly-varying magnetic field. The simulations are performed with a variety of parameters and boundary conditions: in low, high and of-order-unity plasma-$\beta$ regimes, with periodic and fixed boundaries, with and without a friction force, at various resolutions and with various diffusivities. Current sheets form, over a dynamical timescale, under {\it all} conditions explored.  At higher resolution they are thinner, and there is a greater number of weaker current sheets. The magnetic field eventually relaxes into a smooth minimum energy state, the energy of which depends on the magnetic helicity, as well as on the nature of the boundaries.}
\end{abstract} 
 \begin{keywords} ({\it magnetohydrodynamics}) MHD -- pulsars: general -- turbulence -- ISM: kinematics and dynamics -- ISM: structure -- ISM: magnetic fields \end{keywords}

\section{Introduction}\label{intro}

In this introduction pulsar scintillations, coronal heating, and some other relevant astrophysical contexts are reviewed. Afterwards in sections \ref{topology} and \ref{sims} I describe how we might expect magnetic discontinuities to form and present some simulations demonstrating this, before summarising in section \ref{discuss}.

\subsection{Pulsar scintillations}

Pulsar scintillations, a time variability observed in essentially all pulsars, were discovered shortly after the discovery of pulsars themselves \citep{Scheuer:1968}. They are produced by variations in the refractive index of the interstellar medium (ISM), which in turn depends on the density of free electrons $n_e$.  \citet{Lithwick:2001} proposed that the density variations are the result of Kolomogorov turbulence, but this picture is apparently inconsistent with recent observations. \citet{Stinebring:2001} found parabolic shapes in dynamic spectra when viewed in Fourier space, which are best interpreted as being produced by a {\it small number} of scattering screens in the line of sight \citep{Walker:2004,Hill:2005,Cordes:2006,Brisken:2010}.

 To produce the observed time delays we need path-length differences corresponding to scattering angles of order 1-100 mas. The scattering angle is proportional to the density contrast $\Delta n_e$. If the scattering comes from sheets  aligned at arbitrary angles to the line of sight, then we require a contrast $\Delta n_e \sim 10^3 n_e$, which seems difficult to produce by any obvious physical process. Fortunately though, Snell's law tells us that if the angle of incidence is small (i.e. grazing incidence) then the scattering angle is inversely proportional to the angle of incidence. Consequently we can have $\Delta n_e \sim n_e$ as long as the angles of incidence are of order $10^{-3}$. This is much easier to produce with plausible physical processes.

Given that any mechanism to produce sheet-like structures is likely to produce them throughout the ISM, but that observations require only one or a few scattering structures at any one time, this grazing incidence picture has the advantage that if the required grazing angle is $10^{-3}$, only a fraction $10^{-6}$ of sheets in the line of sight will produce appreciable scattering. To best explain observations, they should be of order $0.1$ AU in thickness, separated by perhaps $0.1$ pc. An alternative geometry of filaments, as seen in cosmological dark matter simulations, not only requires alignment in two directions rather than one direction, but it is also physically difficult to explain, as self-gravitating objects for instance.

\citet{Goldreich:2006} proposed that these structures are current sheets. \citet{PenLevin2014} constructed a model where the scintillations are produced by density variations associated with `ripples' -- magnetic waves propagating along current sheets. This can reproduce the shapes seen in the secondary dynamic spectra. 

\subsection{The solar corona}\label{corona}

Current sheets are thought to play a major role in coronal heating (\citealt{Parker:1972,Parker:2012} \& refs.\ therein; Galsgaard \& Nordlund 1996). As relevant to the topic of this paper, there are two differences between the corona and the ISM: while the ISM has a plasma-$\beta$ of order unity\footnote{The plasma-$\beta$ is defined as the ratio of gas pressure to magnetic pressure: $\beta\equiv 8\pi P/B^2$.}, the solar corona has a low-$\beta$ and its magnetic field is subject to the force-free condition $\curl {\bf B}\approx\alpha{\bf B}$. The other difference is the boundary: motion in the corona is driven by convective motion at the boundary, i.e. the photosphere. This motion has a timescale significantly longer than the dynamic timescale in the corona, so the coronal magnetic field evolves in quasi-static equilibrium. The ISM, on the other hand, lacks clearly defined boundaries.

In the `magnetic carpet' picture of coronal heating, convective motion at the photosphere moves the footpoints of coronal magnetic field arches around in uncorrelation, random directions. Without topological reconnection, field lines become tangled around each other, with a complexity increasing with the passage of time. Parker (1972) argued that there is generally no new {\it smooth} force-free equilibrium solution for the magnetic field to relax into, so the new equilibrium must contain tangential discontinuities, which are presumably sites of reconnection and heating. Although this reasoning does not apply close to a boundary with imposed smooth conditions, it should become asymptotically valid at large distance from the boundary. At a large distance from the boundary we are in a similar situation to the ISM -- this argument could conceivably also apply to fluids without meaningful boundaries.

{\mkb A topic of debate is whether, when starting with a potential (i.e. current free) coronal field, it is necessary to move the footpoints a certain distance before discontinuities appear. If discontinuities appear as soon as the field is moved even slightly away from a potential configuration, and the discontinuities allow the field to relax efficiently back into a potential state, then the coronal field would be in a quasi-statically evolving near-potential state. This would be a serious problem for coronal heating, because moving the footpoints of a potential field requires no work, and no energy could be transferred from the convective motion into the corona. If however the discontinuities do not form immediately, or if the discontinuities do not immediately produce rapid reconnection, then the coronal field would instead be in a quasi-static non-potential force-free equilibrium, presumably always slightly over the threshold for the appearance of current sheets (unless some instability threshold is crossed, as happens in flares), and the moving footpoints would experience a magnetic drag force.}

\citet{Galsgaard_Nordlund:1996} performed simulations in which motion at the photosphere results in the formation of discontinuities in the corona, at which reconnection takes place. If the boundary driving is halted, the reconnection apparently stops but the discontinuities remain; reconnection at the discontinuities resumes once the driving is switched back on again. It would therefore be possible for an equilibrium to contain discontinuities, but perhaps only with certain boundary conditions.

Various authors have disputed this model. For instance \citet{CraigSneyd2005} find in numerical experiments with boundary displacements that smooth equilibria can form. However, \citet{Low2013} points out a boundary problem with their method of frictional relaxation, leading to false solutions; damping via fluid viscosity may be more physical than a friction force. 

\subsection{Other contexts}

{\mk Massive stars may also have magnetically active coronae. Stars above about 7 $M_\odot$ have a convective layer just below the surface resulting from an iron-ionisation heat-capacity bump (see \citealt{Cantiello:2009}).} This layer could host a dynamo, and the field generated should easily reach the surface via buoyancy \citep{Cantiello:2011}. {\mk These stars emit X rays, but with current observational data it is not possible to ascertain whether the source of X rays is close to the stellar surface, in which case there is probably solar-type magnetic heating, or further away, in which case it could be shock heating associated with} the line deshadowing instability in the wind. However, various kinds of variability do point to activity close to or on the stellar surface (\citealt{Michaux2014} and refs.\ therein).

{\mk There is evidence of discontinuities in the solar wind, interpreted by \cite{Borovsky2008} as flux tubes and by \citet{Li2008} as current sheets. In an effort to explain the latter, \citet{Zhdankin2012,Zhdankin2013} performed simulations in reduced MHD, using an incompressible equation of state and a strong guide field. They find discontinuities, and have some success in matching them statistically to observations. \citet{Greco2010} also find discontinuities in incompressible simulations. Discontinuities were apparently already found in similar simulations by \citet{MuellerBiskamp2000}; unfortunately they are impossible for the reader to see because the results are presented only Fourier space.}

The Voyager probes recently flew through the reverse shock of the solar wind {\mk (termination shock)}, and it appears that Voyager 1 may now be flying through the contact discontinuity {\mk (heliopause)} into the (perhaps shocked) ISM (e.g. \citealt{Schwadron:2013}). Both probes have flown through numerous current sheets. They may be the manifestation of the rotation of the Sun combined with the Sun's magnetic field being tilted with respect to the rotation axis, but the irregular intervals between discontinuities speak instead in favour of magnetic reconnection events (\citealt{Opher:2011}).

{\mk Underdense X-ray bubbles in galaxy clusters, inflated by magnetised AGN jets, often emit sychrotron radiation from cosmic rays. Often a considerable distance from the central AGN, the cosmic rays must presumably be produced {\it in situ}, perhaps in magnetic reconnection sites. The presence of these would not be surprising:} as \citet{Braithwaite:2010} and \citet{Gourgouliatos2010} have argued, the magnetic field in a bubble should undergo plenty of reconnection over a long timescale.

\section{Formation of discontinuities in MHD}\label{topology}

In this section the formation of current sheets is examined from the point of view of magnetic topology, using the ISM as an example context. 
The ISM is stirred amongst other things by supernovae, by stellar winds and ionising radiation, by the galactic disc's self-gravity, shear flow and associated instabilities. These sources drive motions on a variety of scales, and most of them are {\mk transient}; in addition, the dynamic (acoustic or Alfv\'enic) timescale is {\mk often smaller at the stirring length scale than the intervals between stirring events.}   It is natural to expect therefore that {\mk at least} {\it locally} the ISM relaxes towards some equilibrium state which evolves quasi-statically in response to changing external conditions. This state would persist until it is destroyed by some event.

Note that the formation of current sheets was apparently mentioned by \citet{BiskampBook}, as part of a discussion of turbulent cascades.

\subsection{Topological relaxation}

Locally one can consider what should happen after a stirring event. Initially, the terms present in the momentum equation are unbalanced. The momentum equation and the sizes of the various terms are:
\beqa\label{eq:mom}
\frac{{\rm d}{\bf u}}{{\rm d}t} \;\;=\;\; - \frac{1}{\rho}{\bm\nabla}P \;\;+\;\; \frac{1}{4\pi\rho}({\bm\nabla}\times{\bf B})\times{\bf B}\;\;+\;\;\nu\nabla^2{\bf u}\\
U^2 \;\;\;\;\;\;\;\;\;\;\; \frac{\Delta\rho}{\rho}c_{\rm s}^2  \;\;\;\;\;\;\;\;\;\;\;\;\;\;\;\;\;\;\;\;\; v_{\rm A}^2 \;\;\;\;\;\;\;\;\;\;\;\;\;\;\;\;\;\;\;\;\; \frac{\nu U}{L}\;\;\nonumber
\eeqa
where ${\bf u}$, $P$, $\rho$, ${\bf B}$ and $\nu$ are the velocity, pressure, density, magnetic field and viscosity, where $U$ and $L$ are typical flow speeds and length scales, $\Delta\rho$ is a typical density variation, and $c_{\rm s}$ and $v_{\rm A}$ are typical sound and Alfv\'en speeds. To get the second line, I have multiplied by $L$ and assumed that $U\sim L/T$ where $T$ is a typical timescale. In the ISM we have approximate equipartition between kinetic, thermal and magnetic energies, and the first three terms in the momentum equation are all comparable in size. {\mk The last term, whose size $\nu U/L$ can be written as $U^2/{\rm Re}$ where Re is the Reynolds number, is enormously smaller than the other terms in almost all astrophysical contexts (one exception being the gas in galaxy clusters; see e.g. \citealt{Shakeachicken2006}). This leads inevitably to nonlinear dynamics.}

 In the absence of driving the system will head towards an energy minimum. Kinetic and magnetic energy is dissipated into thermal energy via shocks and reconnection as well as by diffusion, at least on the smaller scales. As can be seen by comparing the sizes of terms in (\ref{eq:mom}), this will happen on the dynamic timescale. The end-state is then an equilibrium where the pressure gradient and Lorentz forces balance each other:
\beq\label{eq:gen}
\del P=\frac{1}{4\pi}(\curl {\bf B})\times{\bf B}.
\eeq
The Lorentz force is perpendicular to the magnetic field, and in equilibrium must also be perpendicular to the isobars, meaning that field lines reside on surfaces of constant pressure. Picking an arbitrary starting point and following a field line, one visits every point on a surface of constant pressure -- one speaks of a surface-filling magnetic field. A general vector field, however, is volume-filling, meaning that following a field line one eventually visits every point in space. Generally, one would expect that when a stirring event has finished, the ISM will be left with a volume-filling magnetic field\footnote{As long, of course, as the stirring also changes the previously established topology, which there is no reason to dispute since one expects reconnection during stirring as well as during relaxation.}.  In almost all astrophysical contexts the Ohmic timescale is very long, and flux-freezing should hold to a high degree. How then is it possible to reach the rather special surface-filling topology required by equilibrium? The answer presumably involves small length scales, on which magnetic diffusivity can have a significant effect on a dynamic timescale, resulting in topological reconnection. In other words, spontaneous formation of discontinuities, or current sheets. Their thickness presumably depends on magnetic diffusivity, but this need not affect the global behaviour, just as the viscosity of a gas affects the thickness of a shock but has no effect on conditions on either side.

Note that the situation is different in two dimensions, where there is no topological barrier to equilibrium: in order to satisfy $\dv{\bf B}$ every field line necessarily joins up with itself, and all that is required to reach an equilibrium is to equalise pressure along each line. 
Interestingly, \citet{Gruzinov:2009} (using periodic boundaries) demonstrated that this equilibrium generally contains discontinuities, even though the initial conditions are smooth. In three dimensions, \citet{Arnold:1986} suggests that in the case of infinite conductivity, the equilibrium should consist entirely of discontinuities. Obviously with high but finite conductivity something else happens: perhaps a slowly-evolving equilibrium forms, containing discontinuities separated by smoothly-varying regions, or perhaps an equilibrium forms with no discontinuities at all. The outcome could well depend on the boundary conditions. 

\subsection{Magnetic helicity}\label{sec:hel}

Magnetic helicity is a global quantity defined as the volume integral of the scalar product of the magnetic field with its vector potential: $H\equiv(1/8\pi)\int\!{\bf A}\!\cdot\!{\bf B}\,{\rm d}V$. In the case of infinite conductivity, the helicity of a volume is conserved as long as no magnetic flux passes through the boundaries \citep{Woltjer:1958}. It a topological property of the field which cannot be changed by moving the fluid around while the flux is frozen into it. It has units of energy $\times$ length and so is present (in a hand-wavy sense) more in the larger structures than is the energy -- and it seems, at least in many situations, to be {\it approximately} conserved when there is a small but finite magnetic diffusivity, even whilst magnetic energy is being dissipated on small scales. This property has been demonstrated in many contexts from the laboratory \citep{Hsu_Bellan:2002} to the solar corona \citep{Zhang_Low:2003}. {\mk In section~\ref{sec:diff} it can be seen that helicity conservation improves with decreasing magnetic diffusivity.}  Consideration of dimensions gives us
\begin{equation}\label{eq:hel_val}
|H| = l E 
\end{equation}
where $l$ is the some length scale and $E$ is the magnetic energy. In a simple configuration, $l$ should be comparable to the size of system. For a given value of helicity the energy is lowest for a greater $l$ and so equilibria should preferentially form with a high $l$ and a simple, large-scale structure. However there may in principle be local energy minima with smaller $l$ and greater $E$.

\section{Numerical simulations}\label{sims}

We are interested in how an arbitrary smoothly-varying magnetic field evolves in the absence of external driving and relaxes towards equilibrium. The setup here is a cube of side $L$ containing gas with an ideal-gas equation of state and ratio of heat capacities $5/3$. Except for the simulations described in section~\ref{boundaries}, boundaries are periodic. In the initial conditions, we have uniform pressure and density, zero velocity, and a magnetic field produced from a Fourier transformation of random phases in three-dimensional wavenumber space, at a range of wavenumbers from $k=2\pi/L$ (i.e. the smallest possible wavenumber) to $6\pi/L$, with a $k^{-3}$ power law. The detail of this is not important; what is important is that this is a volume-filling magnetic field, which will  require topological reconnection to reach a surface-filling equilibrium, and that the field contains only relatively large length scales.

The code used is the {\sc Stagger Code} \citep{Nor_Gal:1995,Gud_Nor:2005}, a Cartesian finite-difference MHD code with fifth-  and sixth-order spatial derivatives and interpolations and third-order time-stepping. It has a hyperdiffusive scheme which damps structure near the spatial Nyquist frequency whilst preserving structure on larger scales. The hyperdiffusion is switched on for most of the simulations, but switched off for a few, where a slightly more quantitative examination of the effect of diffusion is requried.

In the next section I describe a set of simulations differing only in their random initialisation. In subsequent sections I describe additional simulations with various different parameters.

\subsection{Plasma-${\bm\beta}$ of order unity}\label{sec:fid}
 
\begin{figure*}
\includegraphics[width=0.33\hsize,angle=0]{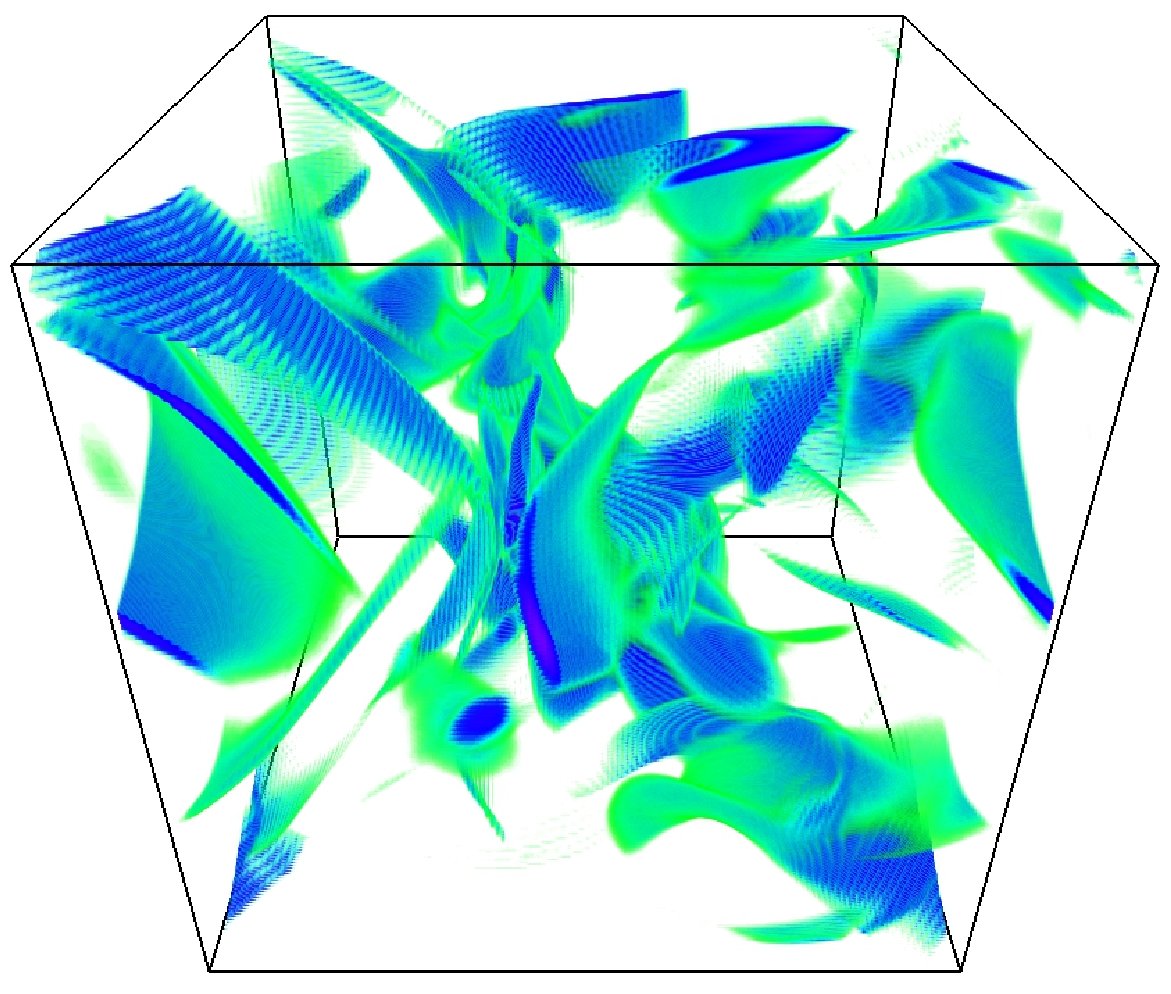}
\includegraphics[width=0.33\hsize,angle=0]{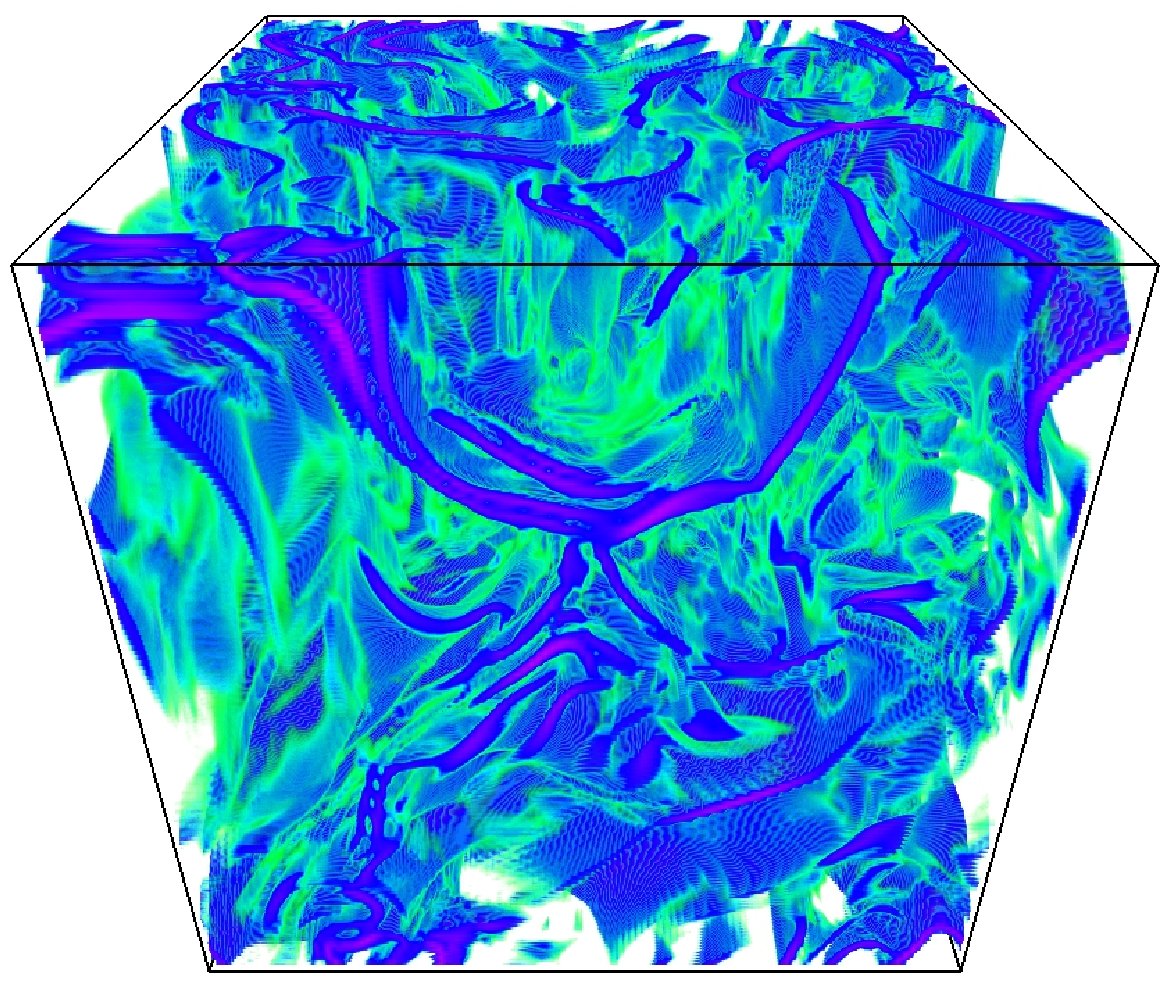}
\includegraphics[width=0.33\hsize,angle=0]{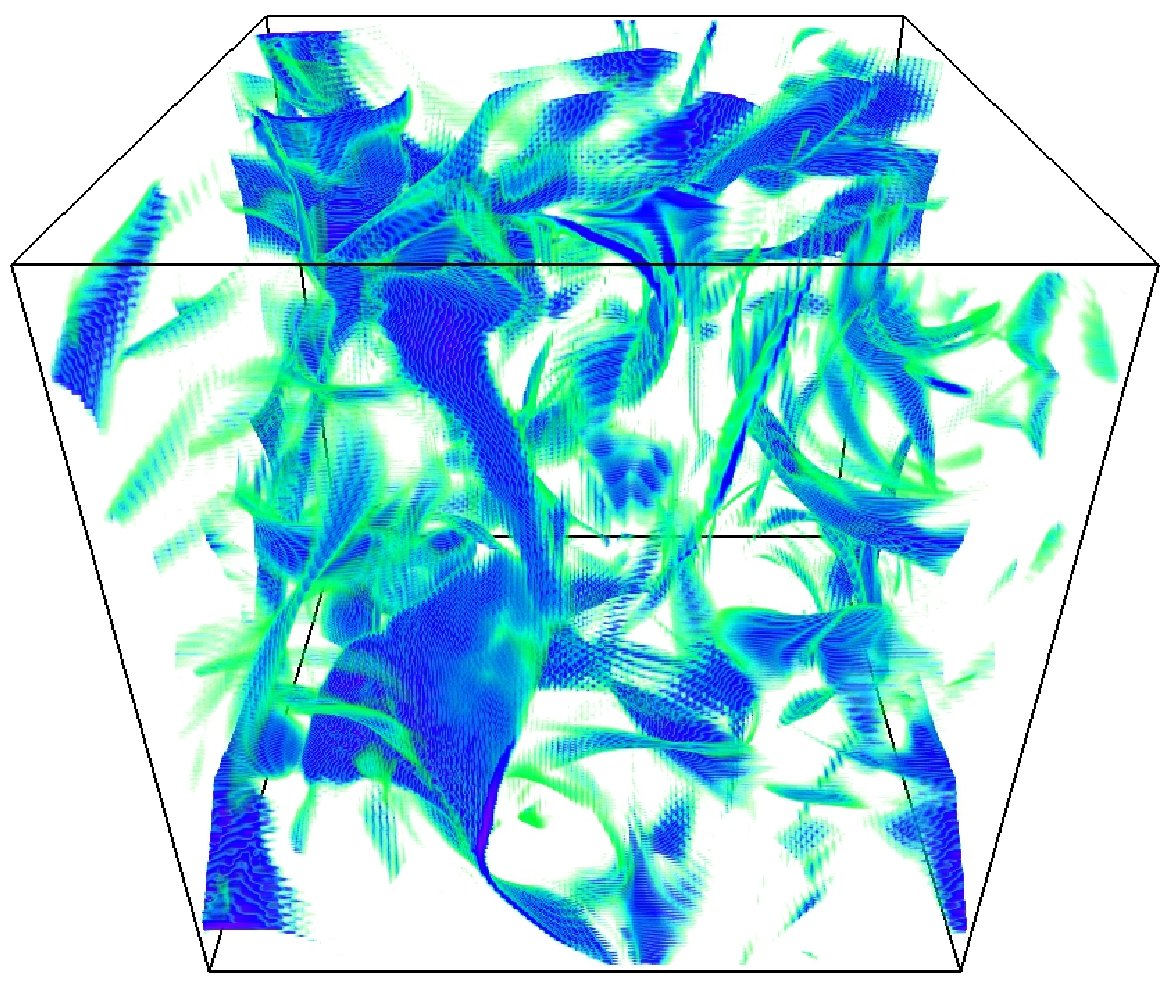}
\caption{Snapshots at $t=1.2$, $4.3$ and $18.3$, from left to right, of the simulation marked 2 in fig.~\ref{fig:fid_EH}. The volume rendering represents current density, with the same thresholds in all  plots: green, blue and purple represent medium, high and very high current density. Current sheets form on the dynamic timescale. The second snapshot is at the time of maximum activity, after which the number and intensity of current sheets drops as equilibrium is approached. The simulations was run at resolution $128^3$ with Pr$_{\rm m}=10$ and hyperdiffusion, and began with a mean plasma-$\beta$ of $1/2$. Note that the mottling effect on the current sheets comes from the visualisation procedure.}
\label{fig:fid}
\end{figure*}
\begin{figure*}
\includegraphics[width=0.33\hsize,angle=0]{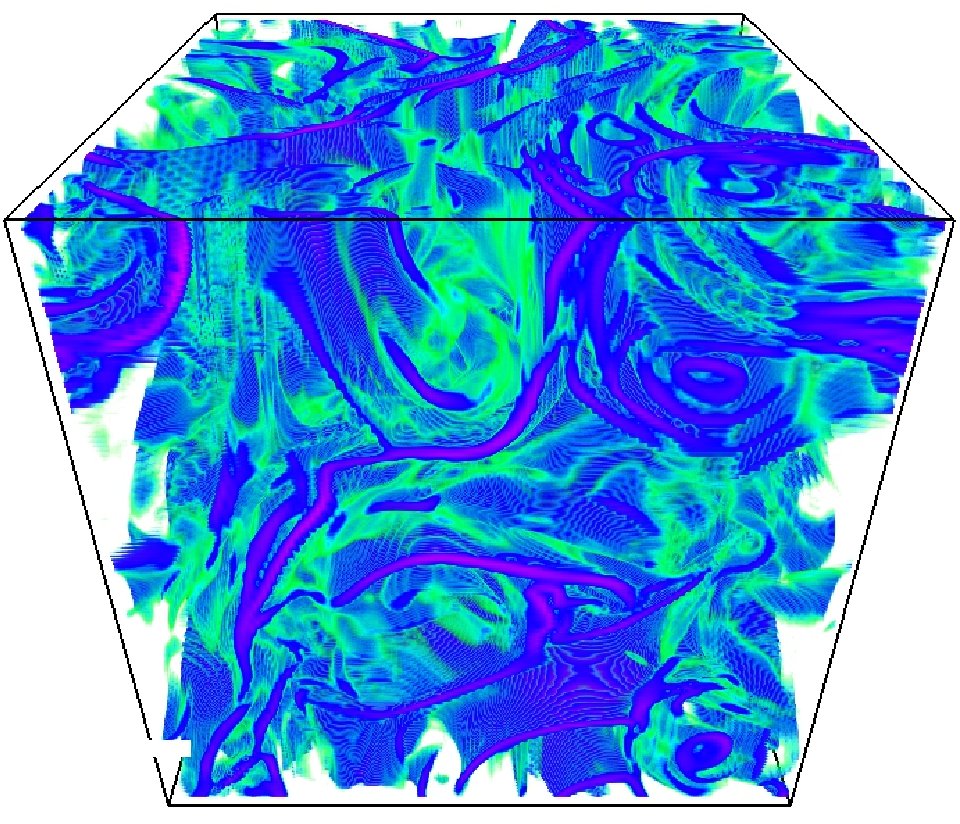}
\includegraphics[width=0.33\hsize,angle=0]{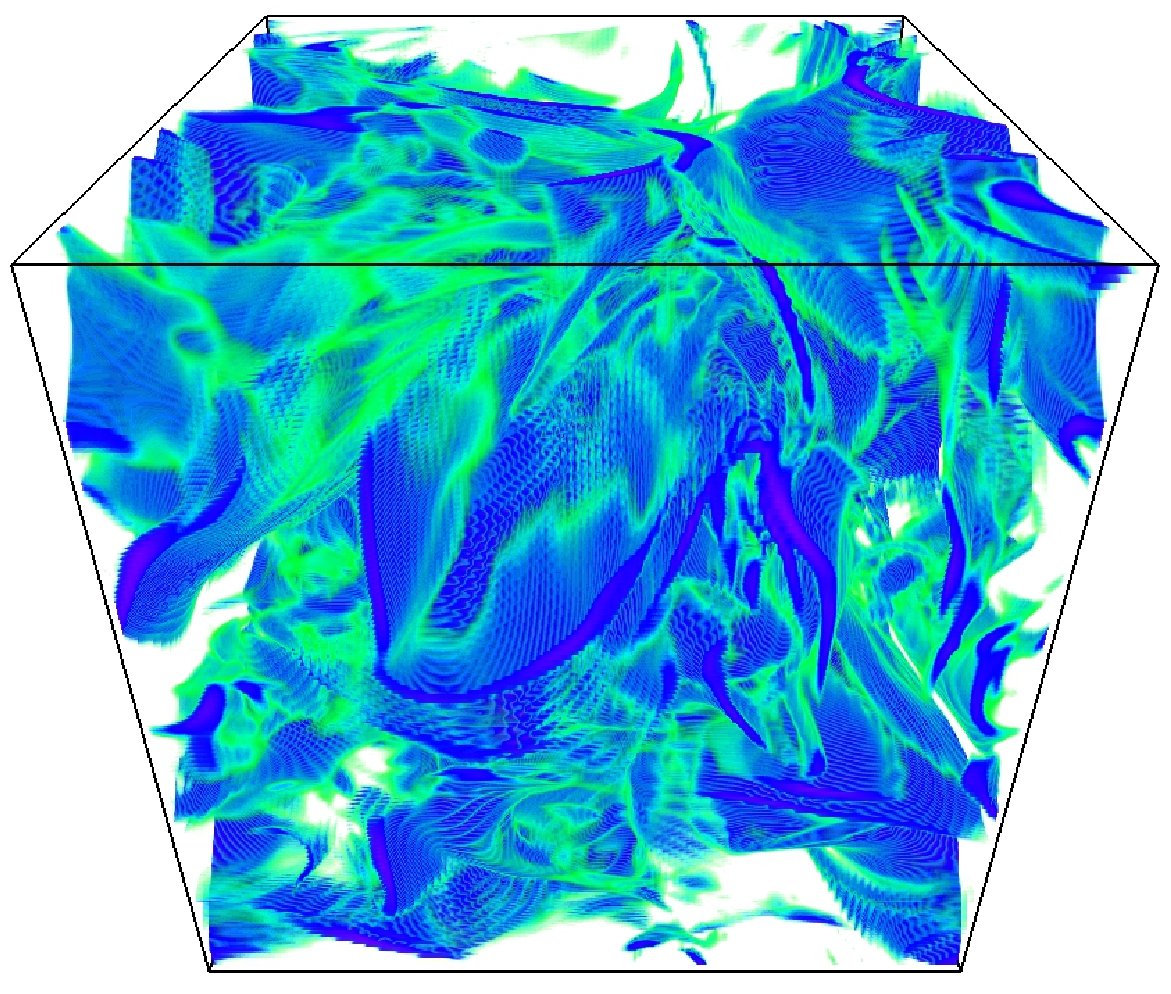}
\includegraphics[width=0.33\hsize,angle=0]{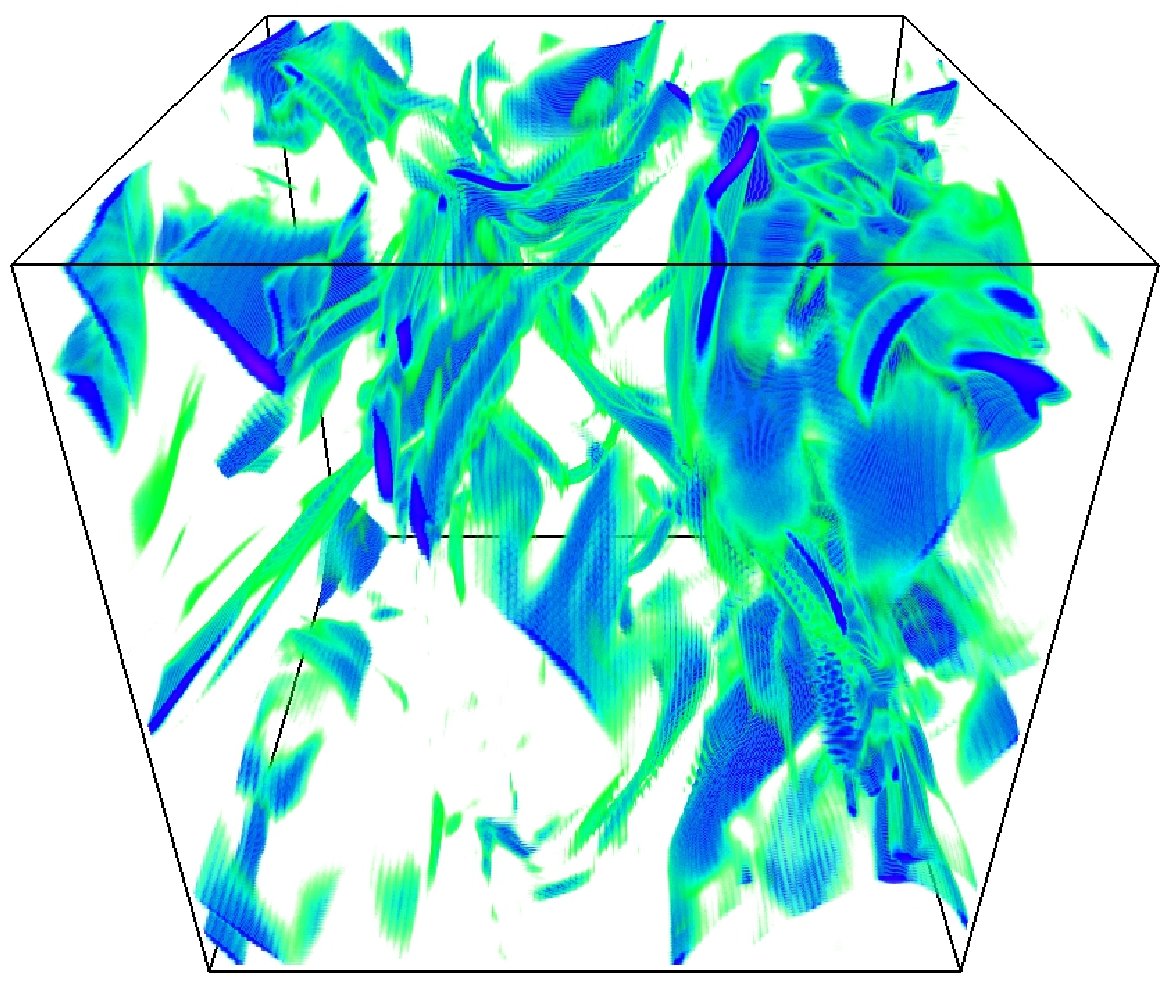}
\caption{Snapshots of three simulations identical to that in fig.~\ref{fig:fid} except for the random seed used to generate the initial conditions; they are marked 1, 3 and 4 in fig.~\ref{fig:fid_EH}. The snapshots are taken in each case approximately at the time of maximum activity: at $t=2.0$, $3.3$ and $4.2$, from left to right; the magnetic helicity in the initial conditions increases in that order. The simulation on the left has zero initial helicity. Note the rolled features on the right-hand-side of that snapshot -- these concentric current sheets annihilate each other shortly after appearing.}
\label{fig:fids}
\end{figure*}

In this section, I present results from simulations run at a resolution of $128^3$ with hyperdiffusion switched on. The magnetic Prandtl number Pr$_{\rm m}\equiv\nu/\eta=10$ (where $\nu$ and $\eta$ are the viscosity and magnetic diffusivity), the reason being that the ISM, solar corona and essentially all other low-density astrophysical environments have high Pr$_{\rm m}$ and 10 is computationally easily achievable. The initial strength of the field is set such that the plasma-$\beta$ has a mean value of $1/2$, which, as we shall see below, results in an equilibrium with $\beta\sim 1$, as in the ISM. Several simulations were performed with these parameters, with different random seeds used to create the initial conditions. Note that the time units used in this paper are normalised such that an Alfv\'en wave takes on average $1/2$ time units to travel a distance $L/2\pi$ in the initial conditions; this time becomes somewhat longer as the field decays. In other words, the time unit is roughly equal to an Alfv\'en timescale.

\begin{figure}
\includegraphics[width=1.0\hsize,angle=0]{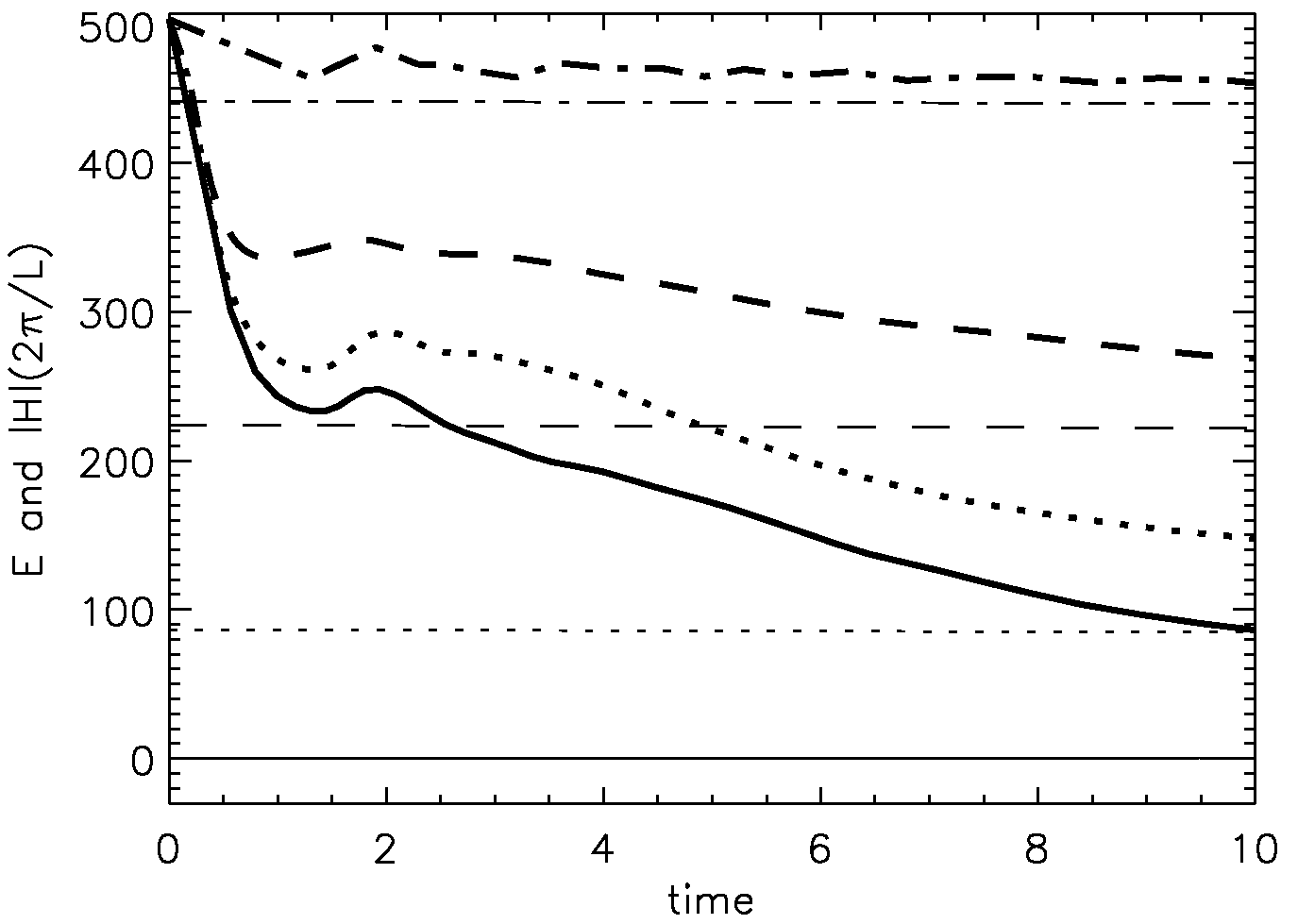}
\includegraphics[width=1.0\hsize,angle=0]{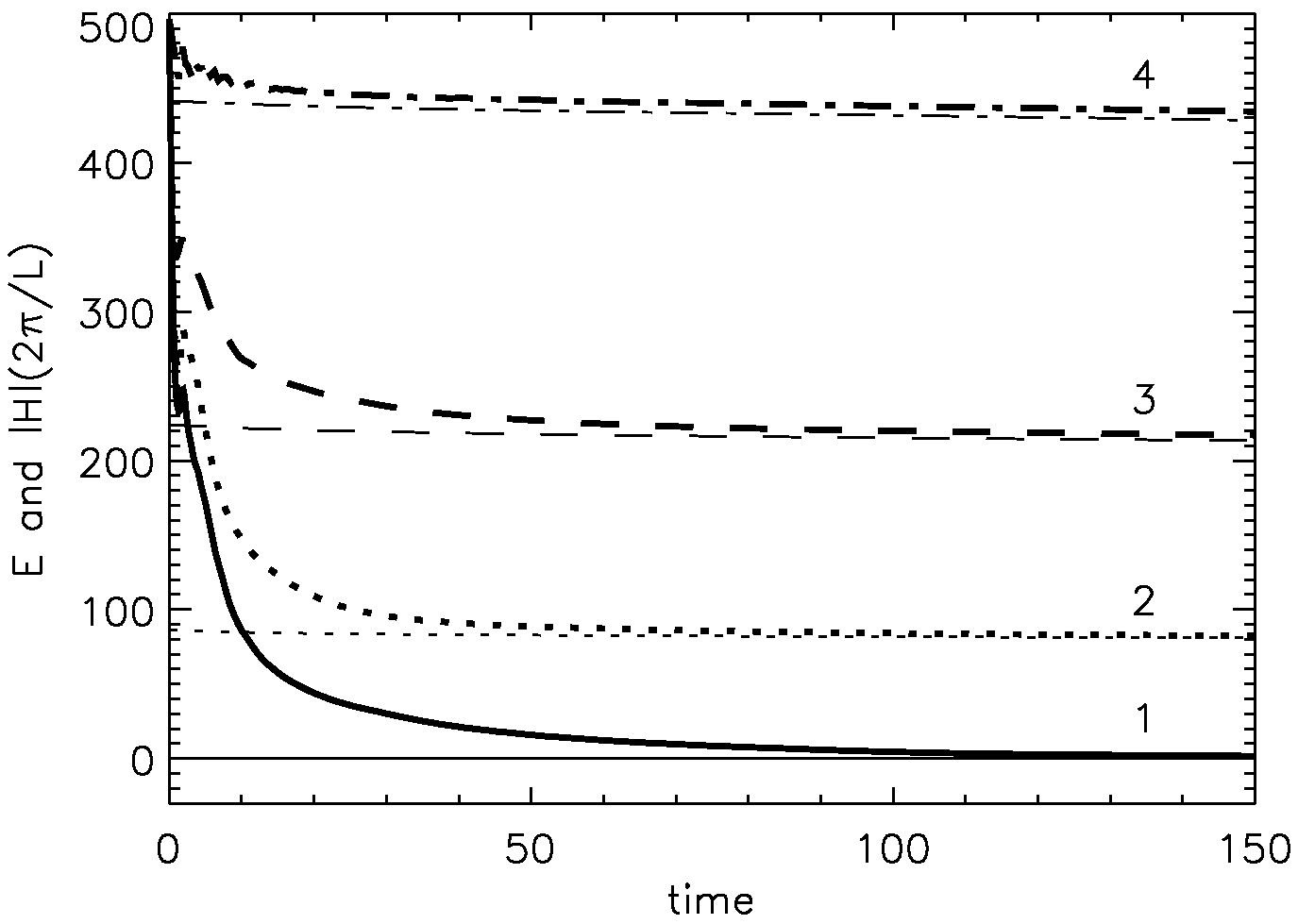}
\caption{The magnetic energy (thick lines) and helicity (thin lines) against time in four simulations, which differ only in their random initialisations, each plotted with a different line style. The upper panel follows the initial evolution, and the lower panel goes to later times. Helicity has units of energy times length so actually plotted on these axes is helicity divided by the length scale $L/2\pi$. Note how helicity is approximately conserved, and how the energy drops from its initial value of $498$  towards the level set by the initial helicity. These simulations are all run at a resolution of $128$ with Pr$_{\rm m}=10$ and hyperdiffusion. Note that simulation 4 begins with zero helicity.}
\label{fig:fid_EH}
\end{figure}

Fig.\ \ref{fig:fid} shows snapshots from one of these simulations at three points in time, and fig.\ \ref{fig:fids} shows snapshots from three more simulations. Multiple current sheets form in each case; the thickness of these sheets is just a few grid spacings. Following on from the ideas mentioned in section \ref{sec:hel}, it is informative to look at the evolution of magnetic energy and helicity in the simulations; this is shown in fig.\ \ref{fig:fid_EH}. The simulations begin with equal energies but different helicities. Clearly, helicity is approximately conserved, and energy drops to a level given by
\begin{equation}\label{eq:min_en}
E = k_{\rm min}H
\end{equation}
where $k_{\rm min}=2\pi/L$ is the minimum wavenumber in a box of size $L$. Note that one simulation (marked 1 in fig. \ref{fig:fid_EH}) has zero initial helicity; its initial conditions are created by adding together two other magnetic fields in the correct ratio. This simulation displays the most reconnection activity, and the simulation with the greatest helicity (marked 4) displays the least. This is not surprising as the lower the helicity, the more magnetic energy is dissipated. Note that because of the `vanishing force-free field theorem' the equilibrium end state is not force-free. (For the 3-line proof of this theorem see \citealp{Roberts1967}, also reproduced in \citealp{Spruit13b}.)

What is actually happening? At first, since the Lorentz force is not balanced by anything, magnetic energy is converted into kinetic. This happens on the Alfv\'en timescale, as we see in the upper panel of fig.\ \ref{fig:fid_EH}. There is some sloshing around between magnetic, kinetic and thermal energy -- note how the magnetic energy rebounds a little after its initial drop. {\mk The average sonic Mach number, starting from $0$, will peak at somewhat less than unity at this stage.} Tangential discontinuities appear: surfaces parallel to the magnetic field, with a field of different direction and magnitude on either side. Between these discontinuities, the field is smoothly varying. Over a dynamical timescale the number of current sheets increases until the situation becomes surprisingly complex: although only large length-scales are present in the initial conditions, eventually there are only relatively small distances between neighbouring current sheets. Current-sheet activity peaks and thereafter the number and strength of current sheets falls as an equilibrium is approached asymptotically. Note that current sheets appear even in simulation 4, which begins with a helicity equal to 90\% of the maximum, and which is therefore already rather close to an equilibrium. This may be relevant to the debate regarding whether discontinuities only form once the configuration is some distance from its energy minimum. A more thorough investigation of this point is left for the future.
 
The discontinuities allow reconnection to take place on the dynamic timescale, or rather, somewhat more slowly: in fig.\ \ref{fig:fid_EH} we can see that after the initial `sloshing' phase and once the current sheets have formed, the evolution slows down by around an order of magnitude. Such a factor of $10$ between the Alfv\'en speed and the reconnection speed is fairly standard in the literature (e.g. \citealt{Elsner:1984,Ikhsanov:2001}). 

Note that at the beginning of these simulations, since the plasma-$\beta$ is of order unity, there is some sonic motion and weak shocks are present. Consequently, some discontinuities appear of the kind which have magnetic flux passing through them. The majority of discontinuities are however of the tangential kind, and after one or two dynamical timescales have passed, all discontinuities are tangential. In the following section a simulation is presented with a high plasma-$\beta$, from which shocks are obviously absent.

\subsection{The high- and low-${\bm\beta}$ regimes}\label{sec:beta}

\begin{figure*}
\includegraphics[width=0.495\hsize,angle=0]{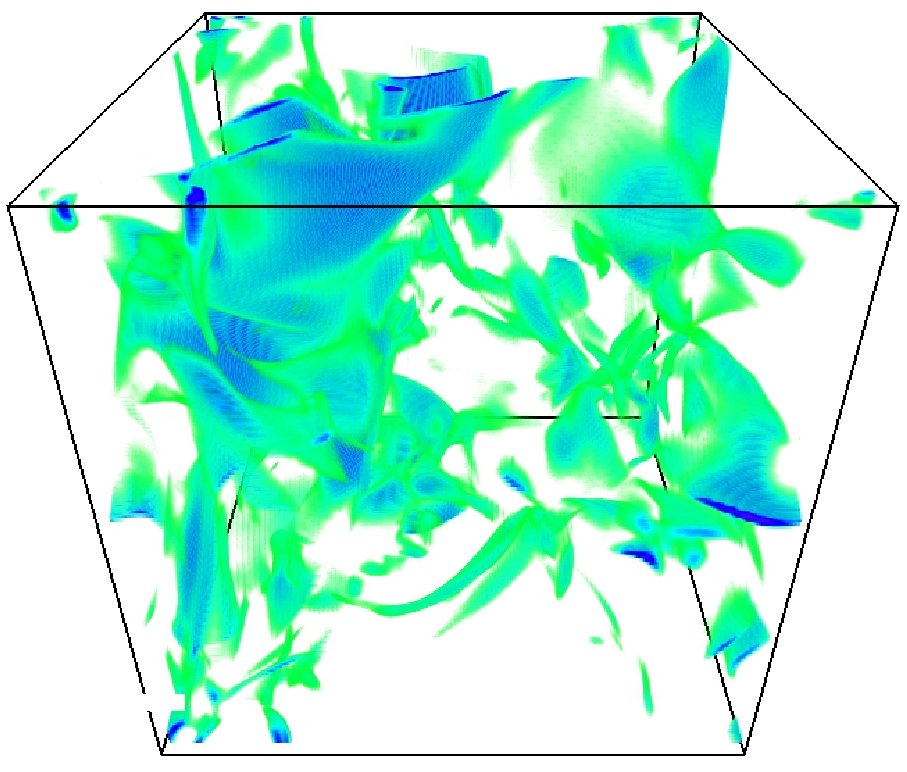}
\includegraphics[width=0.495\hsize,angle=0]{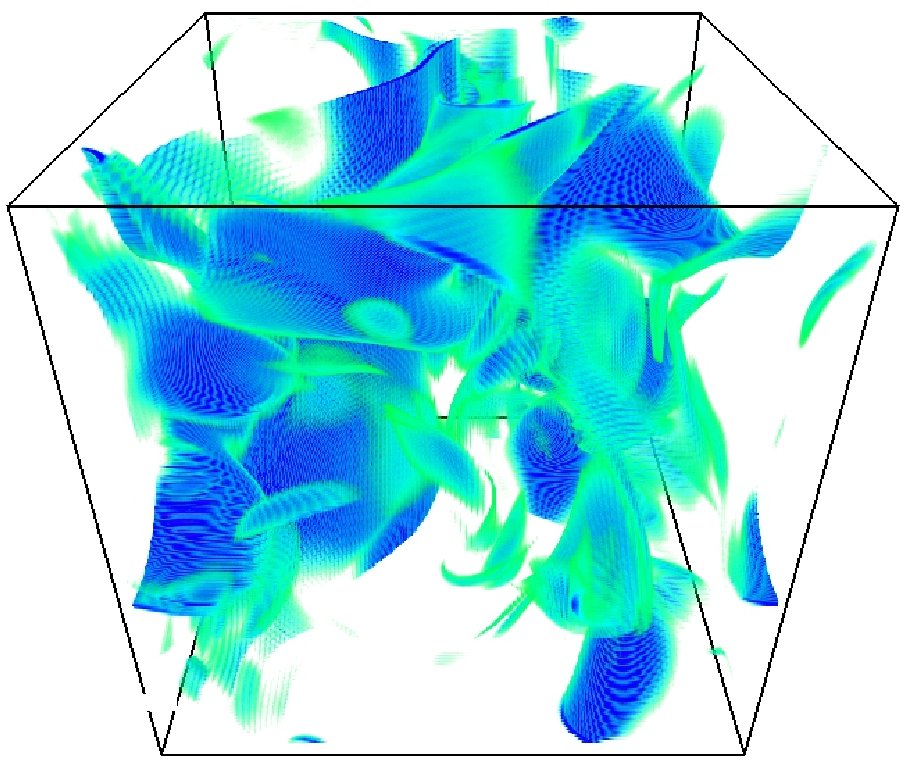}
\caption{Snapshots of two simulations at $t\approx0.6$ and $14.5$, with identical initial conditions except for the mean plasma-$\beta$, which is $1/60$ and $15$ (left and right). The volume rendering represents current density. Clearly, the spontaneous formation of discontinuities is a phenomenon common to both high- and low-$\beta$ regimes.  In both simulations, the spatial resolution is $128^3$, with Pr$_{\rm m}=10$ and hyperdiffusivity switched on.}
\label{fig:beta}
\end{figure*}

The simulations in the previous section had a plasma-$\beta$ of approximately unity, as in the ISM. It would now be interesting to see whether the same behaviour exists in the regimes with high and low plasma-$\beta$; previous results using an incompressible equation of state suggest that current sheets should form in the high-$\beta$ case. To this end, simulations are run with the initial plasma-$\beta$ set to $15$ and $1/60$, which evolve to around $30$ and $1/30$. Ohmic and viscous heating are switched off for the latter, in order to remain in the low-$\beta$ regime\footnote{{\mkb Clearly this is unphysical; real astrophysical plasmas can only remain in the low-$\beta$ regime with the help of physics not present in these simulations, such as radiative cooling, gravity and particular boundaries.}}. Fig.\ \ref{fig:beta} shows snapshots from simulations at high and low $\beta$, using the random initialisation corresponding to simulation 2 in section \ref{sec:fid}. 

The magnetic fields in these simulations have taken different forms, but are qualitatively similar: it is probably safe to conclude that current sheets form in all three regimes. In principle the behaviour might be different at values of $\beta$ more than a factor of 30 away from unity, but there is no particular theoretical reason to expect this.

\subsection{Boundary conditions}\label{boundaries}

\begin{figure*}
\includegraphics[width=0.33\hsize,angle=0]{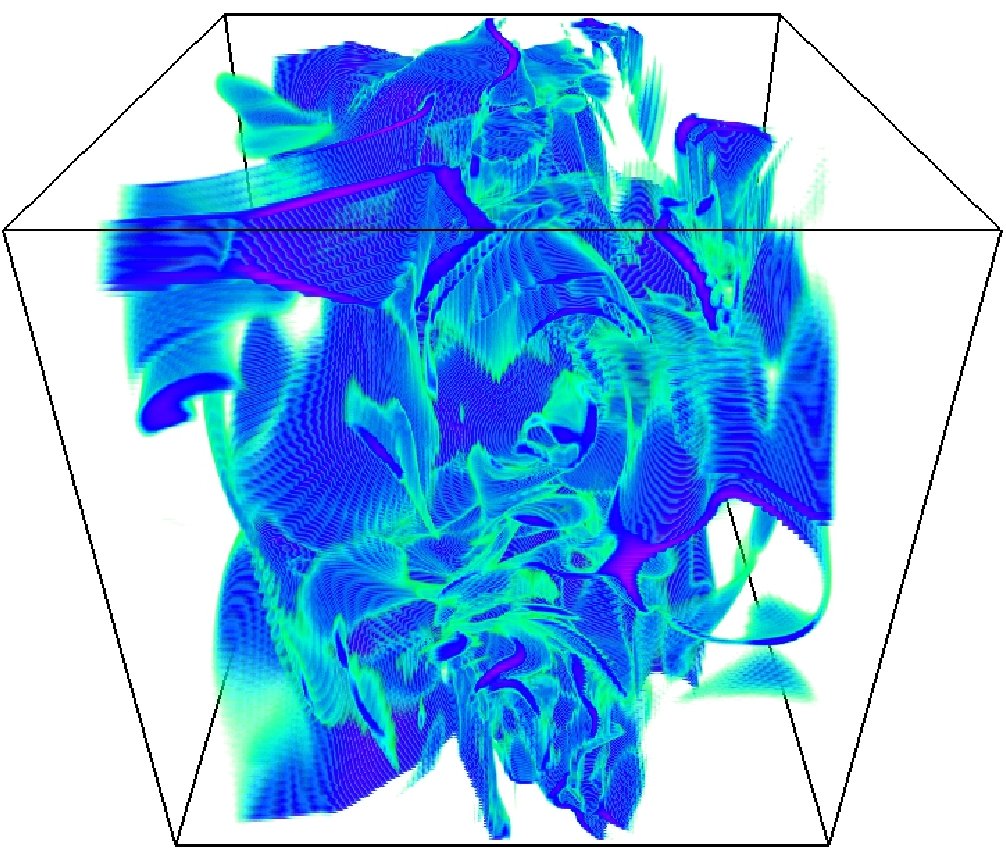}
\includegraphics[width=0.33\hsize,angle=0]{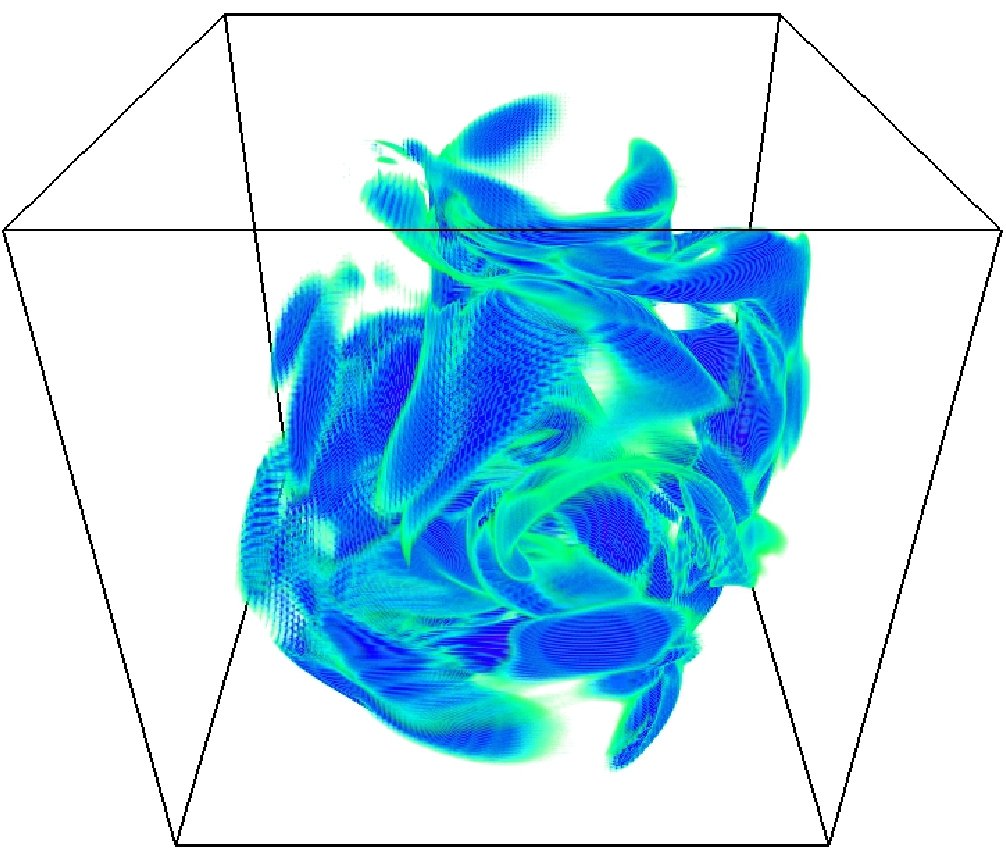}
\includegraphics[width=0.33\hsize,angle=0]{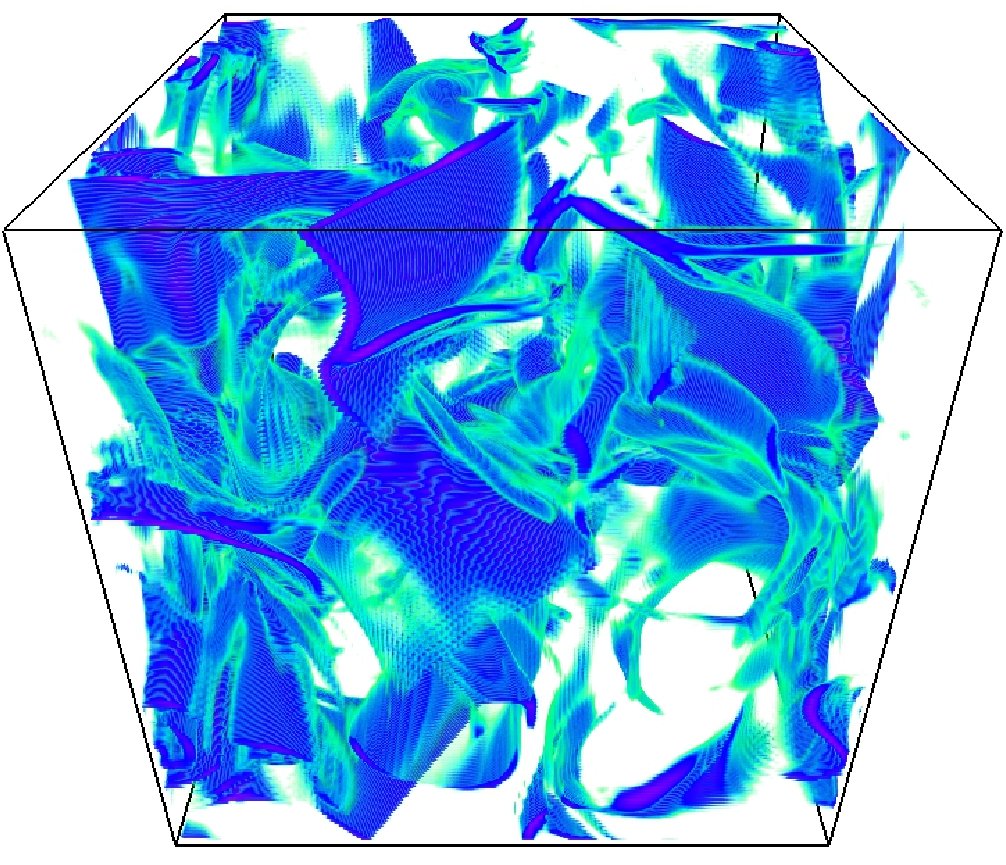}
\caption{{\mk Snapshots from three simulations using the same initial conditions and parameters as simulation 2 from section \ref{sec:fid}, but having different boundary conditions. Left: fixed in one dimension (left and right sides, i.e. at $x=\pm L/2$, as plotted here). Centre: fixed in all three dimensions. Right:  pseudo-vacuum in one dimension (also left and right sides). The snapshots are at times $t=3.3$, $3.6$ and $2.5$.  Current sheets form in all cases, just as with periodic boundaries. All three are run at a resolution of $128$ with Pr$_{\rm m}=10$ and hyperdiffusion, with an average initial plasma-$\beta$ of $1/2$.}}
\label{fig:bounds}
\end{figure*}

\begin{figure}
\includegraphics[width=1.0\hsize,angle=0]{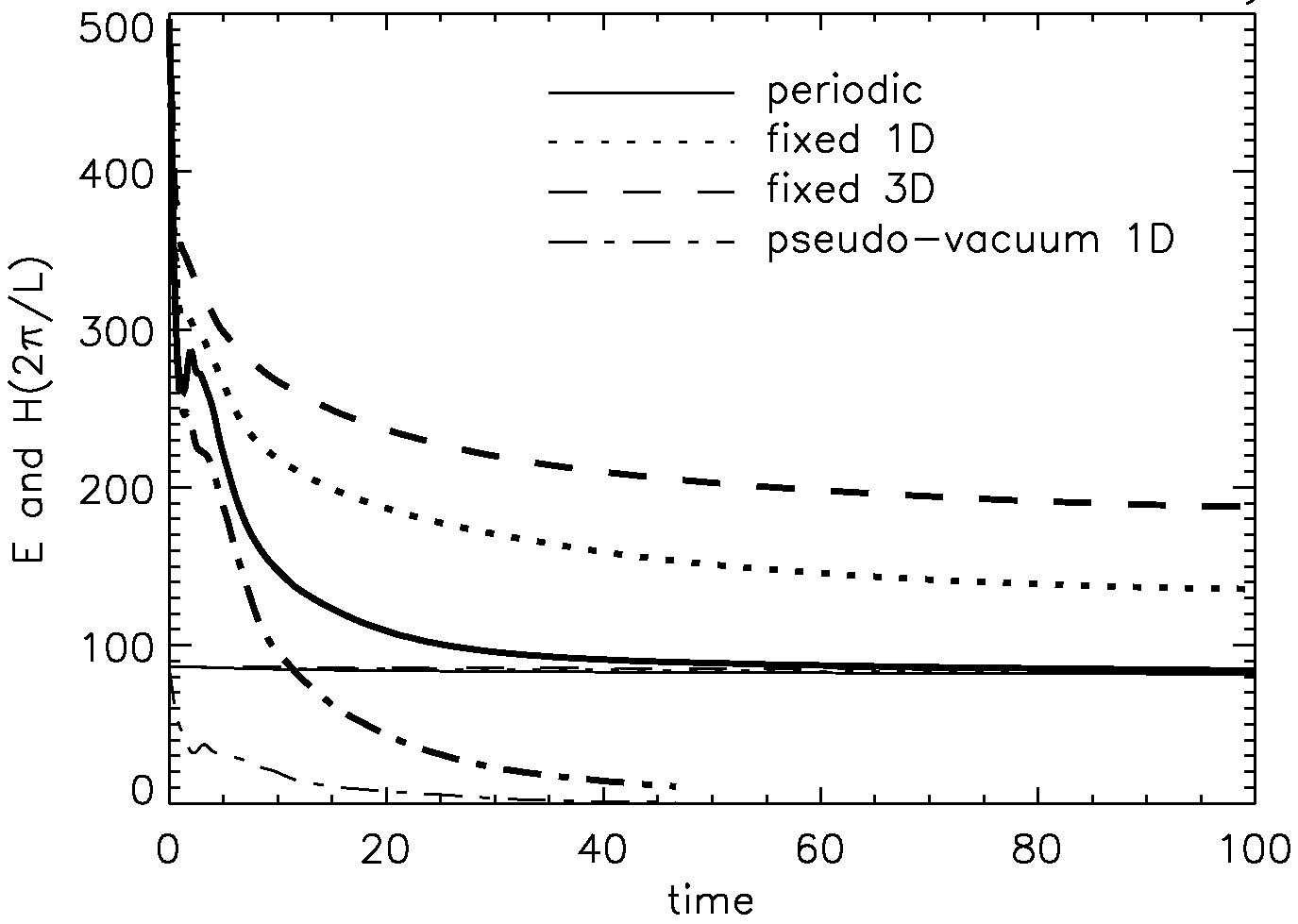}
\caption{{\mk Energy (thick lines) and helicity (thin lines) in four simulations with identical initial conditions but different boundaries. The pseudo-vacuum boundary allows helicity to decay on the dynamical timescale, as expected. The two simulations with fixed boundaries display an approach to equilibrium at constant helicity, but this happens rather more slowly than in the case with periodic boundaries in all three directions. All simulations were run at a resolution of $128$ with Pr$_{\rm m}=10$ and hyperdiffusion, with an initial average plasma-$\beta$ of $1/2$.}}
\label{fig:bounds-helen}
\end{figure}

{\mk The role of boundary conditions in the formation of discontinuities is a topic of some debate, particularly in the context of the solar corona, into which magnetic energy is continually injected via motion at the boundary, i.e. the photosphere. It is therefore a useful exercise to repeat the simulations described above with different boundary conditions. Those most akin to the coronal context would be a fixed boundary in one of the three dimensions, as envisaged in Parker's original model.

In fig. \ref{fig:bounds} we see snapshots from three simulations, all of which use the same initial conditions as simulation 2 from section \ref{sec:fid}. One has a fixed boundary in the $x$ direction, one has fixed boundaries in all directions, and one has a `pseudo-vacuum' boundary in one direction. The latter condition has symmetric $\rho$, $P$, $u_\parallel$ and $B_\perp$, antisymmetric $u_\perp$ and $B_\parallel$, and is essentially a simple way to get similar results to a potential field extrapolation.  In addition, simulations were run with each of these boundary conditions in the high- and low-$\beta$ cases (the same initial conditions as in section \ref{sec:beta}). Current sheets always form, but are in different places in each case; as with periodic boundaries, they appear over a dynamical timescale. The current sheets then become weaker as an equilibrium is approached asymptotically. In the simulations with periodic boundaries, it seems that the equilibrium is smooth (corresponding to minimum energy), but with fixed boundaries it is not completely clear whether the equilibrium contains discontinuities too weak and/or numerous to be seen in the simulations.

In fig.\ \ref{fig:bounds-helen} are plotted the energy and helicity in the three simulations with initial $\beta=1/2$, together with simulation 2 from section \ref{sec:fid}, which has periodic boundaries. {\mkb Fixed boundaries significantly slow the drop in magnetic energy relative to the case with periodic boundaries, and the end state is different, with a higher energy.}  Pseudo-vacuum boundaries do not conserve helicity, as expected. A more detailed study of the role of boundaries is described in a forthcoming paper.}

\subsection{Frictional relaxation}

{\mk In many studies of relaxation into a minimum energy state, the fluid is not free to slosh around, but is instead forced to move directly towards an equilibrium. This is often done by assigning the fluid velocity as the sum of the pressure gradient and Lorentz forces times some multiplier. In other words, the momentum equation (\ref{eq:mom}) is modified: the left hand side changes from $\rd {\bf u}/\rd t$ to ${\bf u}/\tau_{\rm fric}$ where $\tau_{\rm fric}$ is a constant timescale, and the viscous term is dropped. This simplifies the relaxation process; the system finds an equilibrium much faster. However, according to \citet{Low2013}, this method can lead to the wrong equilibrium being reached.

To change fundamentally the nature of the momentum equation within the framework of the MHD code used here would be time consuming, but fortunately there is a simpler solution to mimic this: the addition of a friction force $-\bf{u}/\tau_{\rm fric}$ on the right-hand side of the momentum equation (\ref{eq:mom}). We have something like critical damping if $\tau_{\rm fric}$ is comparable to the dynamic timescale, or in other words if the friction term on the RHS and the inertia term on the LHS of the momentum equation (\ref{eq:mom}) are comparable. If $\tau_{\rm fric}$ is smaller than the dynamic timescale, the fluid is `overdamped' and inertia becomes irrelevant; this is the regime closest to the pure frictional relaxation, we which wish to investigate. Calling the timescale on which the field would otherwise evolve $\tau_0$, which as we saw above is comparable to $\tau_{\rm A}$ or a little longer, then in the regime $\tau_{\rm fric}\ll\tau_{\rm 0}$ we expect the evolution to take place on a timescale $\tau_0^2/\tau_{\rm fric}$. This comes from assuming balance between the Lorentz force and the friction force, as opposed to balance between the Lorentz force and inertia as we had previously. 
   
Simulations were run in the high-, low- and of-order-unity-$\beta$ regimes (as described in section \ref{sec:beta}) with friction force timescales in each case approximately ten times smaller than the Alfv\'en timescale. In fig.\ \ref{fig:f1} we see a simulation otherwise identical to simulation 3 from section \ref{sec:fid}, so with initial plasma-$\beta$ set to $1/2$. There are fewer current sheets, which is presumably because the slower relaxation towards equilibrium means that some current sheets have already come and gone before others form, rather than all forming at the same time. Also, the reduction in reconnection speed increases the thickness of the current sheets, which is not at all surprising.

{\mk We see essentially the same qualitative behaviour: the formation of current sheets on a dynamic timescale, in all three plasma-$\beta$ regimes. Inertia plays no role in these simulations, so we see that the formation of current sheets is essentially a matter of topology. Importantly though, the addition of a friction force does result in a quantitatively completely different end-state, albeit with the same magnetic energy. In some contexts this difference might be quite important.}

\begin{figure}
\includegraphics[width=1.0\hsize,angle=0]{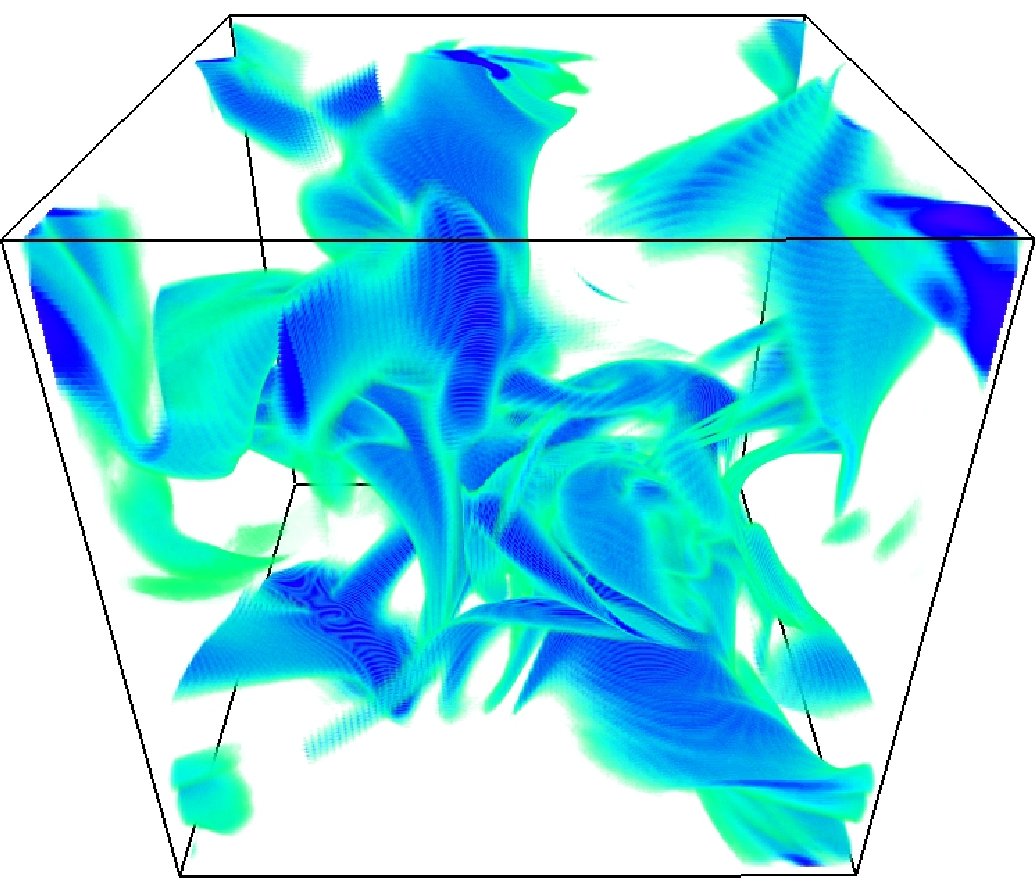}
\caption{A snapshot at time $t=60$ from a simulation with a friction force with a timescale $\tau_{\rm fric}=0.1$, which is about a tenth of the Alfv\'en timescale. Otherwise, this simulation is the same as the simulation 3 from section \ref{sec:fid}. In comparison to that simulation we have here fewer, and somewhat thicker, current sheets.}
\label{fig:f1}
\end{figure}

\subsection{The effects of diffusivity and resolution}\label{sec:diff}

\begin{figure*}
\includegraphics[width=0.33\hsize,angle=0]{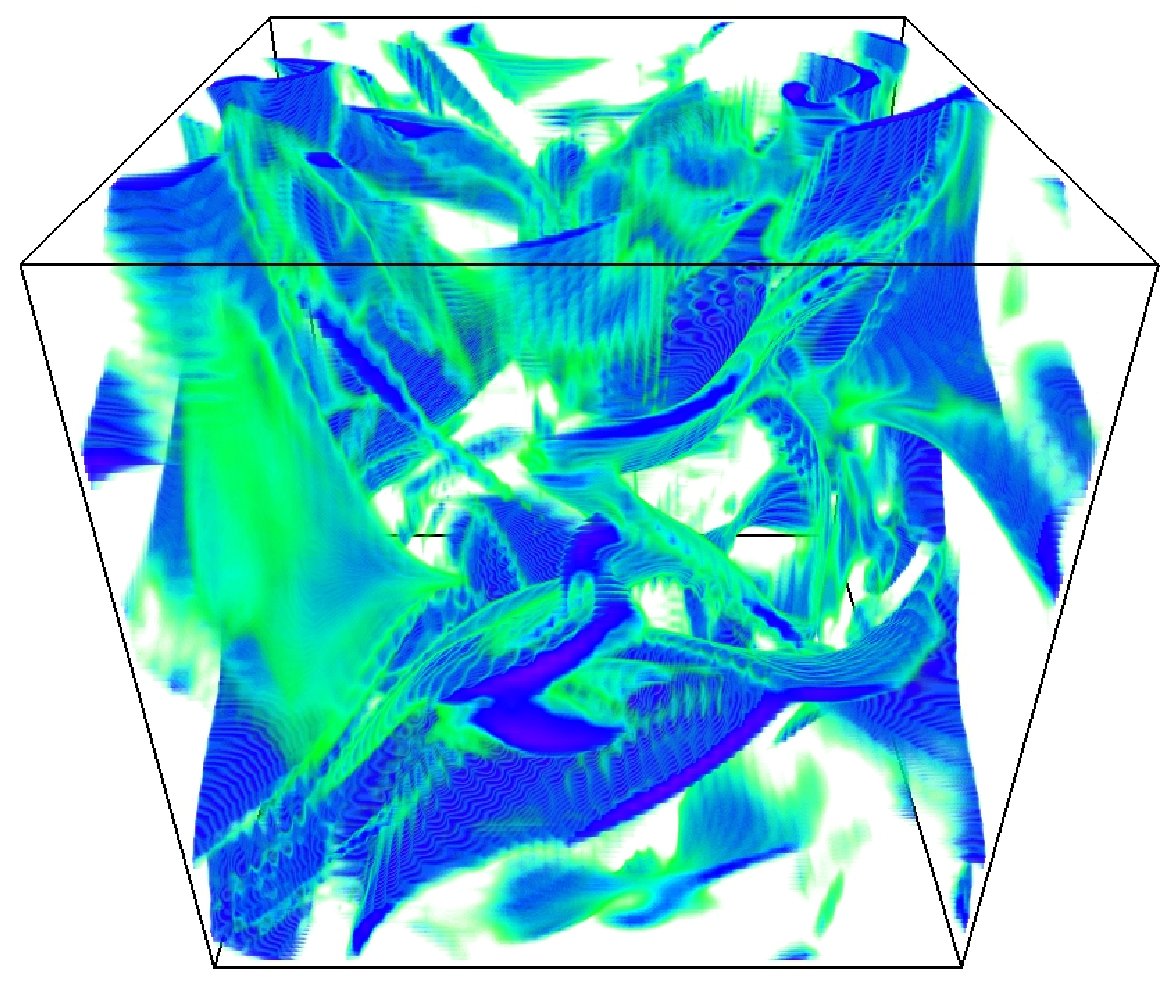}
\includegraphics[width=0.33\hsize,angle=0]{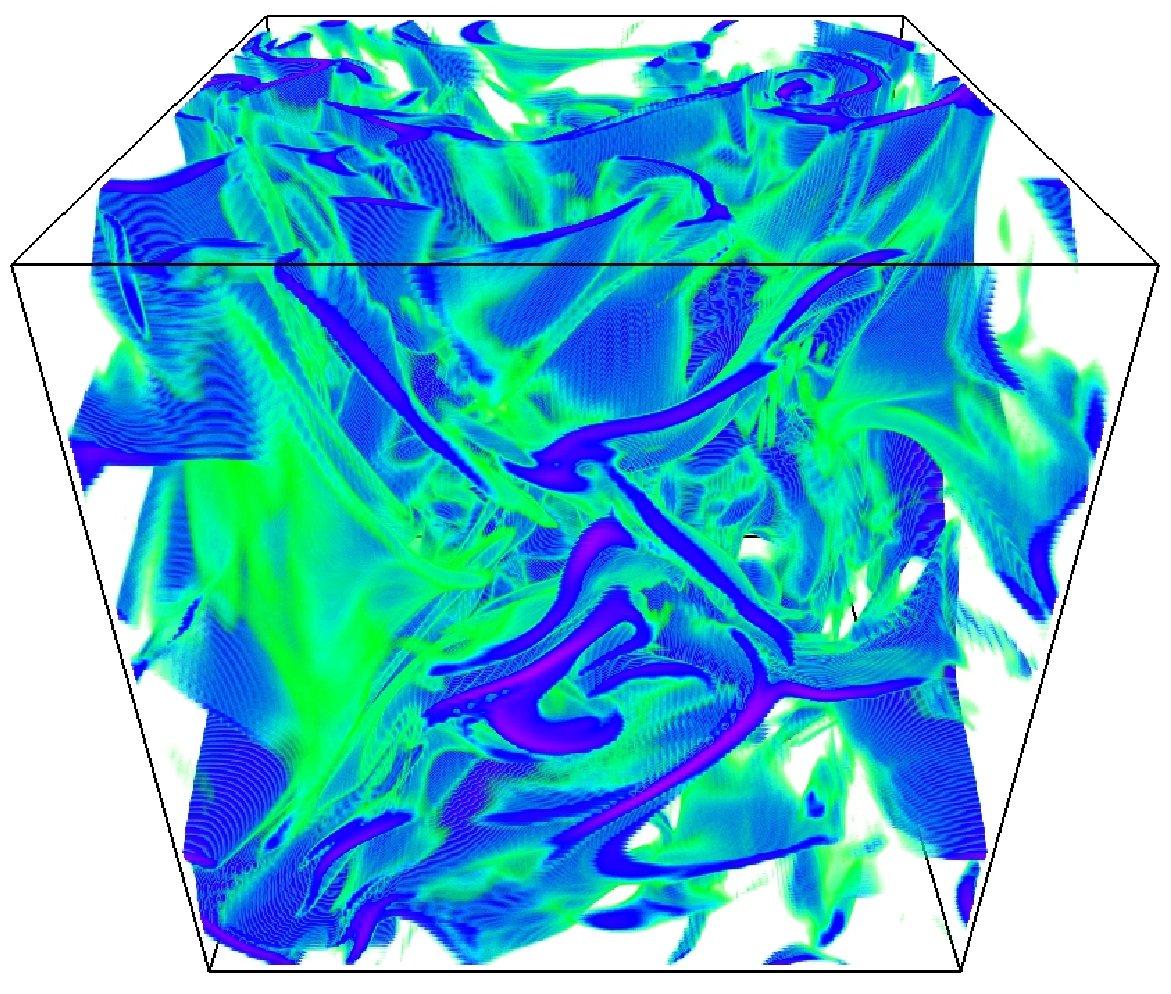}
\includegraphics[width=0.33\hsize,angle=0]{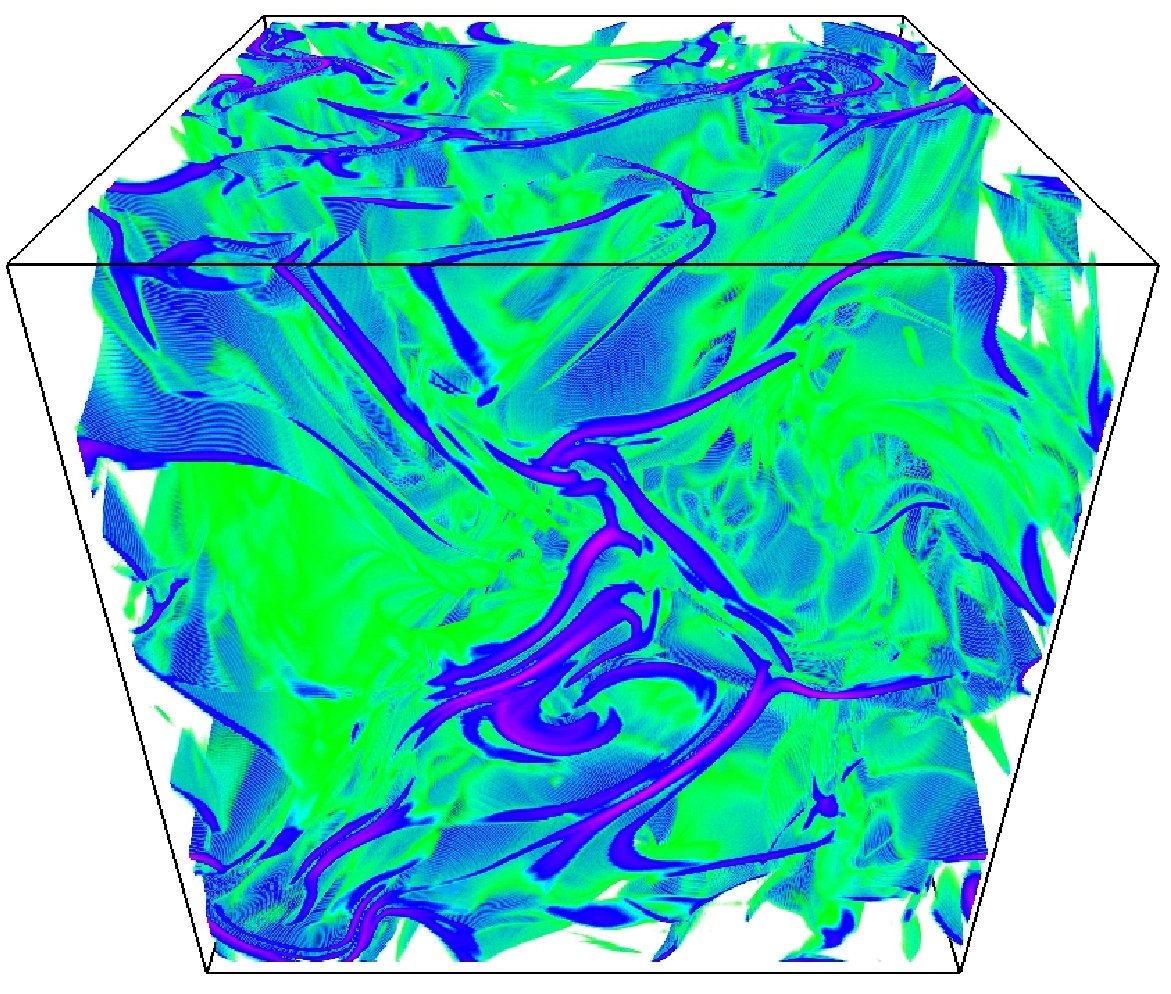}\\
\includegraphics[width=0.33\hsize,angle=0]{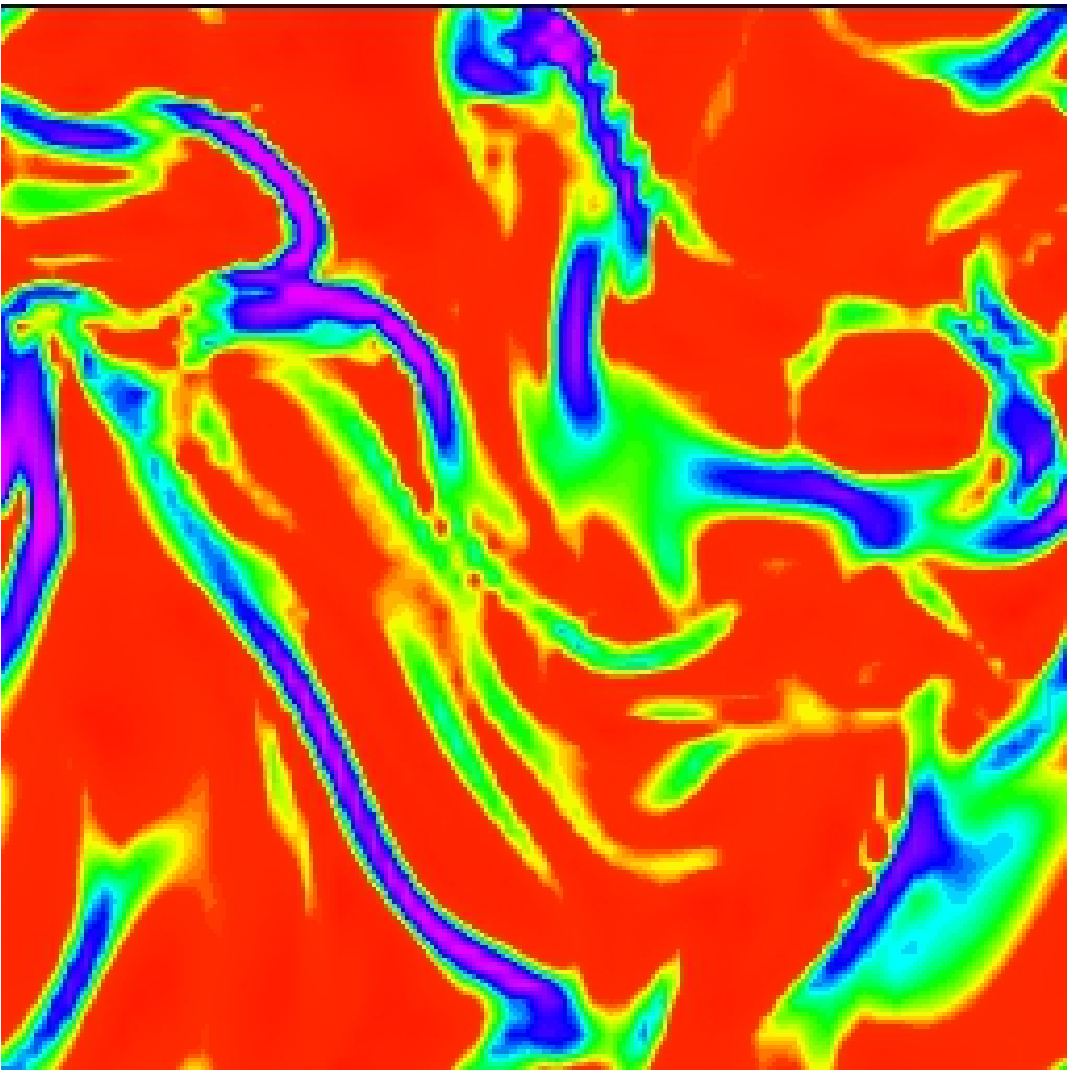}
\includegraphics[width=0.33\hsize,angle=0]{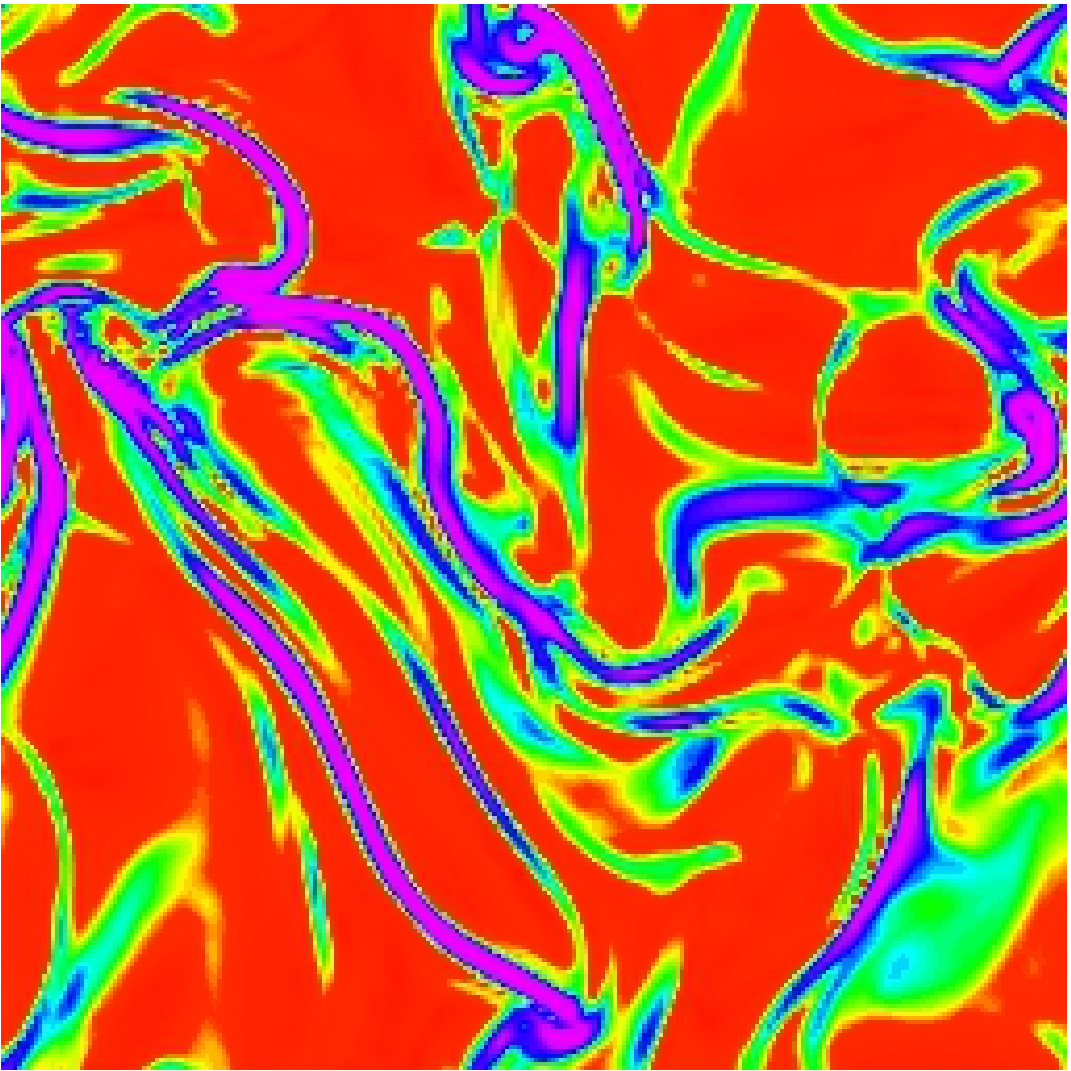}
\includegraphics[width=0.33\hsize,angle=0]{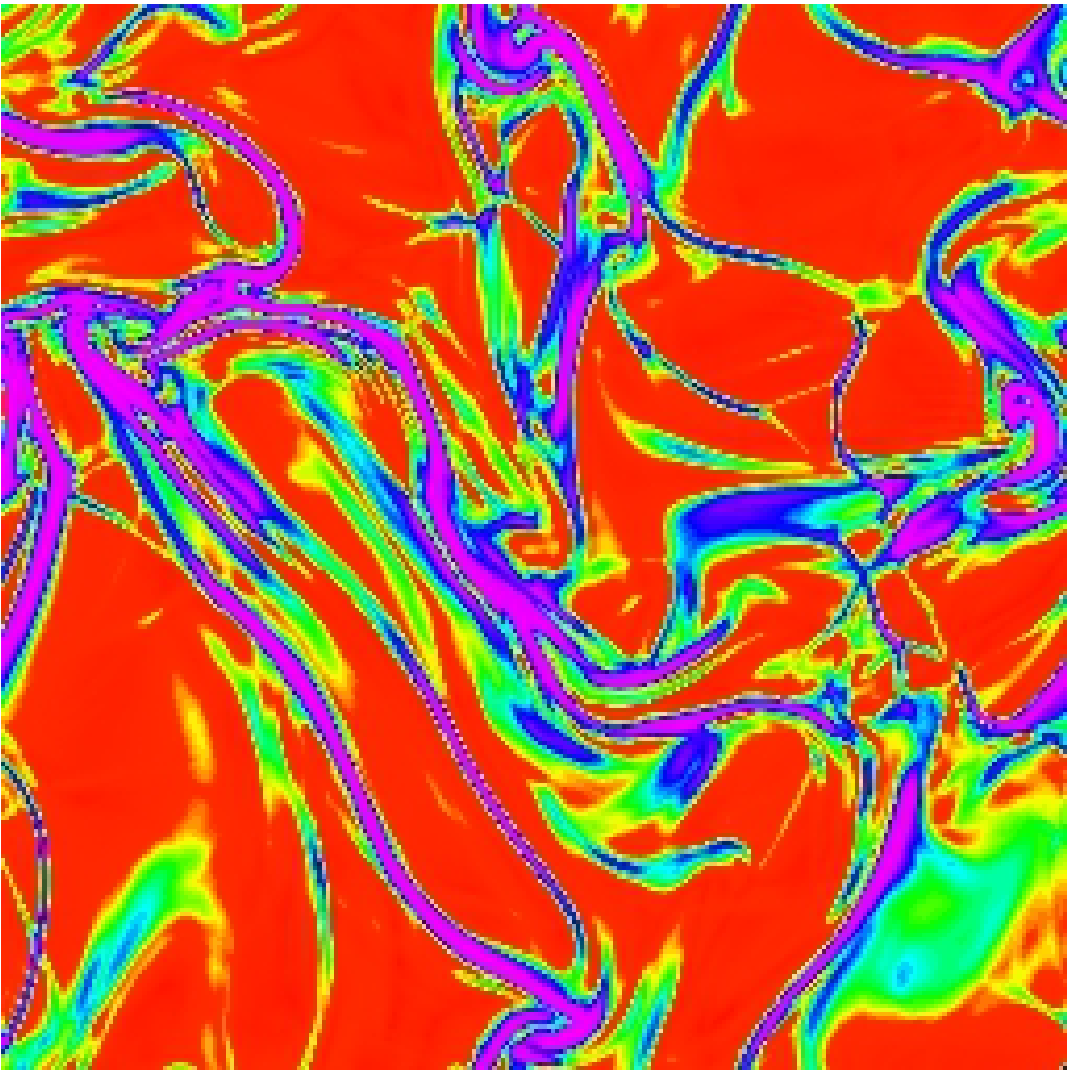}
\caption{Snapshots of three simulations at $t\approx2.5$, with identical initial conditions but with spatial resolutions of $64$, $128$ and $256$ (left to right). Top row: the volume rendering represents current density. Bottom row: slices through the simulation with current density represented by a similar colour-coding as in the 3D plots: red is below-average, then increasing current density is green, blue, then purple. Whilst the larger features are present at all three resolutions, there are additional weaker current sheets at higher resolution. All simulations have an initial mean plasma-$\beta$ of $1/2$.}
\label{fig:res}
\end{figure*}

\begin{figure}
\includegraphics[width=1.0\hsize,angle=0]{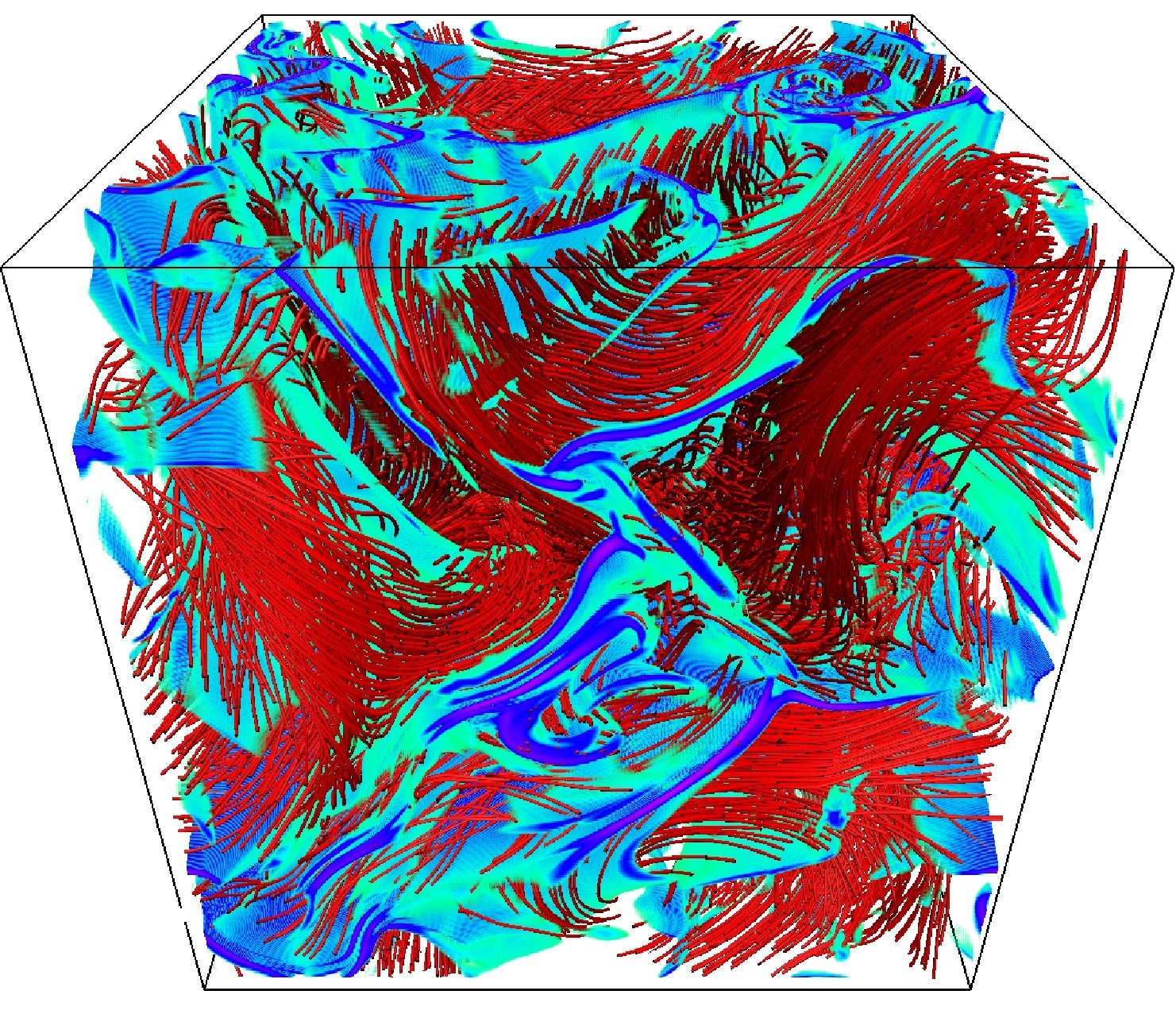}
\caption{Snapshot of the resolution 256 simulation at the same point in time as in fig.\ \ref{fig:res} (right-hand panels). The colour thresholds for current density are slightly different from that figure, and field lines are shown in red.}
\label{fig:res-big}
\end{figure}

A degeneracy is expected here since increasing the resolution is conceptually the same as decreasing the diffusivity. Or to be more precise, increasing the resolution allows a reduction in diffusivity. The simulations presented in previous sections were all run at a resolution of $128^3$, with Pr$_{\rm m}=10$ and hyperdiffusivity switched on, with the minimum magnetic diffusivity required to reliably remove unpleasant zig-zags.

First, we can look at simulations with resolutions of $64^3$ and $256^3$, i.e. with higher and lower effective diffusivities, but which are otherwise identical. Fig.\ \ref{fig:res} shows snapshots from simulations at all three resolutions, using the random initialisation corresponding to simulation 2 from section \ref{sec:fid}. Clearly, the strongest features are visible in all three frames. At higher resolution, the current sheets are somewhat thinner, and there is an abundance of weaker features which are not present at low resolution. Fig.\ \ref{fig:res-big} shows the $256$ resolution run (same snapshot) with field lines. 

\begin{figure*}
\includegraphics[width=0.33\hsize,angle=0]{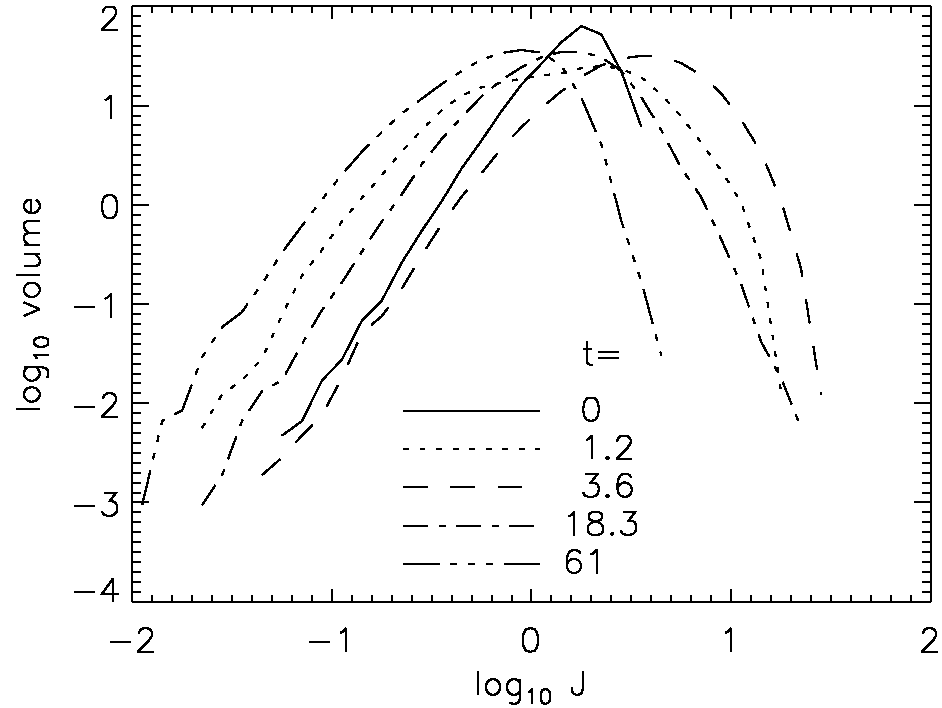}
\includegraphics[width=0.33\hsize,angle=0]{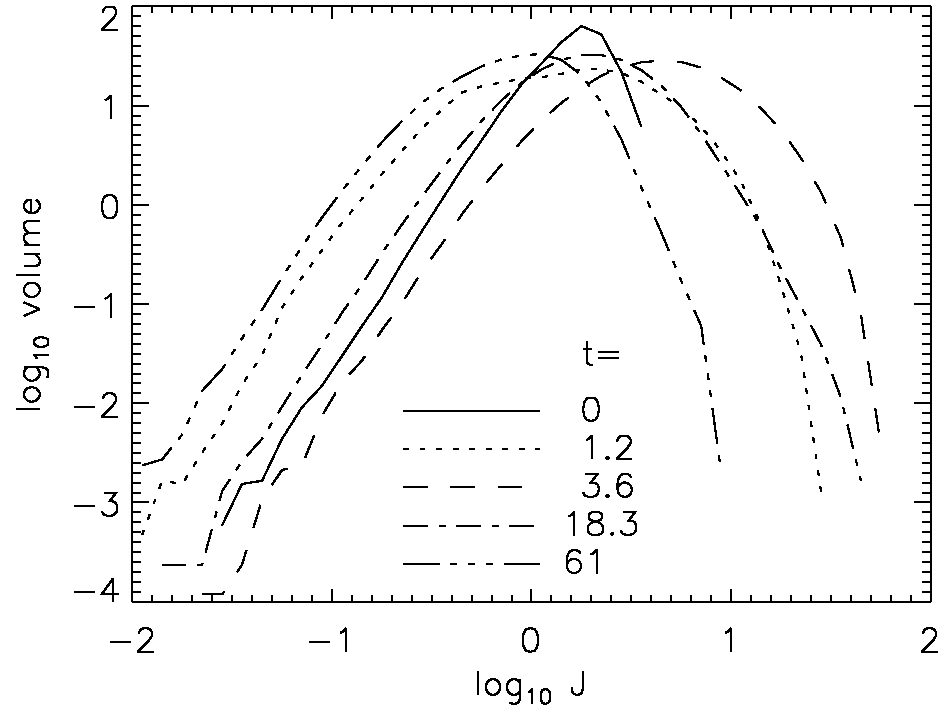}
\includegraphics[width=0.33\hsize,angle=0]{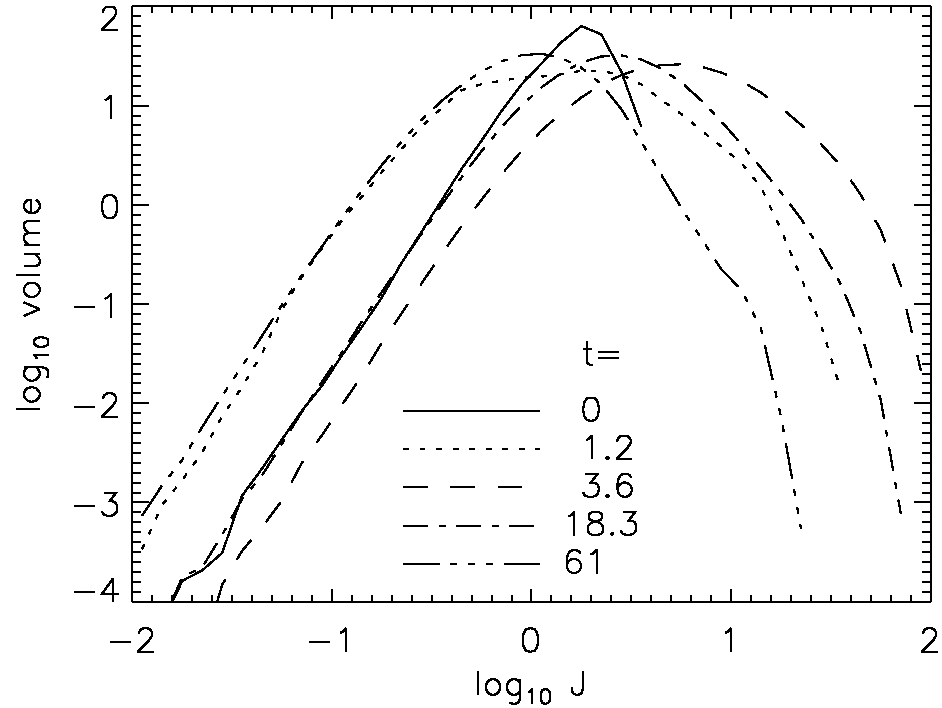}
\caption{Histograms of current density in the three simulations shown in fig.\ \ref{fig:res}: left to right, at resolutions 64, 128 and 256, each at various points in time. In these simulations Pr$_{\rm m}=10$. Note that over the dynamical timescale regions with very high current density develop. These then gradually fade away as the magnetic field approaches its minimum energy state, the current sheets becoming weaker. This effect is more pronounced at higher resolution, where the discontinuities are thinner and the current density correspondingly greater.}
\label{fig:logJ}
\end{figure*}

It is informative to look at the distribution of current density in the simulations. In fig.\ \ref{fig:logJ} histograms of current density are plotted for the three simulations in fig.\ \ref{fig:res} at various points in time. We see that after only a short time, some modest fraction of the volume hosts very high current density. This is the tangential discontinuities. After their formation, they fade in intensity as the field evolves towards its energy minimum -- the magnetic field vectors on either side of a discontinuity gradually approach each other. 

One can also investigate the effect of varying the magnetic diffusivity, or in other words, varying the magnetic Prandtl number while holding the viscosity constant. Simulations were run otherwise identical to simulation 2 from section \ref{sec:fid}, which had  Pr$_{\rm m}$=10, with Pr$_{\rm m}=$3 and 1. To also allow a more quantitative study, three further simuations were run with normal, rather than hyper- diffusion, with the same values of Pr$_{\rm m}$. The effect is essentially the same as that of changing the resolution: lower diffusivity leads to thinner current sheets and more weaker sheets. In fig.\ \ref{fig:helicity} we have at the top a plot of energy and helicity in these six simulations, and below, the timescale of helicity decay, calculated as $\tau_{\rm hel}\equiv (\rd \ln |H|/\rd t)^{-1} $. 

 In these simulations with normal, not hyper- diffusion, helicity conservation begins to fail significantly. Helicity decays -- it never grows -- and it decays faster in simulations with higher magnetic diffusivity. Furthermore, in the simulations with `normal' diffusion, there is a visible quantitative correlation, namely that rate of helicity decay is proportional to magnetic diffusivity. With hyperdiffusion, helicity conservation is better and the same trend is present, albeit with a gentler scaling with diffusivity. This trend seems to hold at all times, right from the beginning when everything is moving around on the Alfv\'en timescale, through the peak of reconnection activity to later times when the current sheets have become much weaker and fewer. At late times the helicity decay timescale, at least in the case with normal diffusion, approaches an asymptotic value equal simply to the global Ohmic timescale.

{\mk Another effect of the magnetic Prandtl number is on {\it how} the magnetic energy is dissipated: we can see in fig.\ \ref{fig:dissipation} that Ohmic dissipation accounts for a greater fraction of total dissipation when Pr$_{\rm m}$ is smaller. In astrophysical contexts with high Pr$_{\rm m}$, such as the ISM where its value is around $10^{11}$, that would mean that almost all of the energy is dissipated viscously.}

\begin{figure}
\includegraphics[width=1.0\hsize,angle=0]{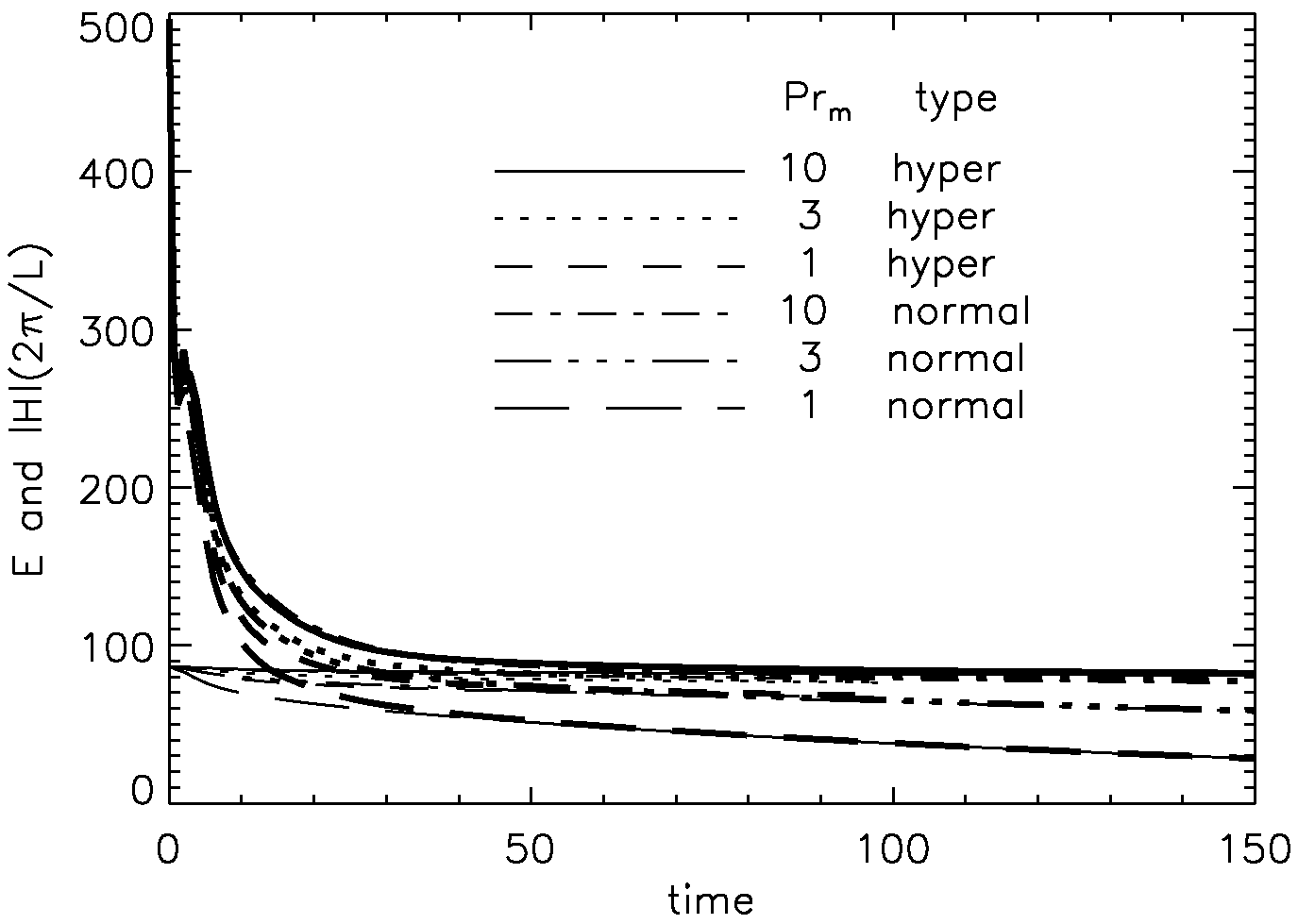}
\includegraphics[width=1.0\hsize,angle=0]{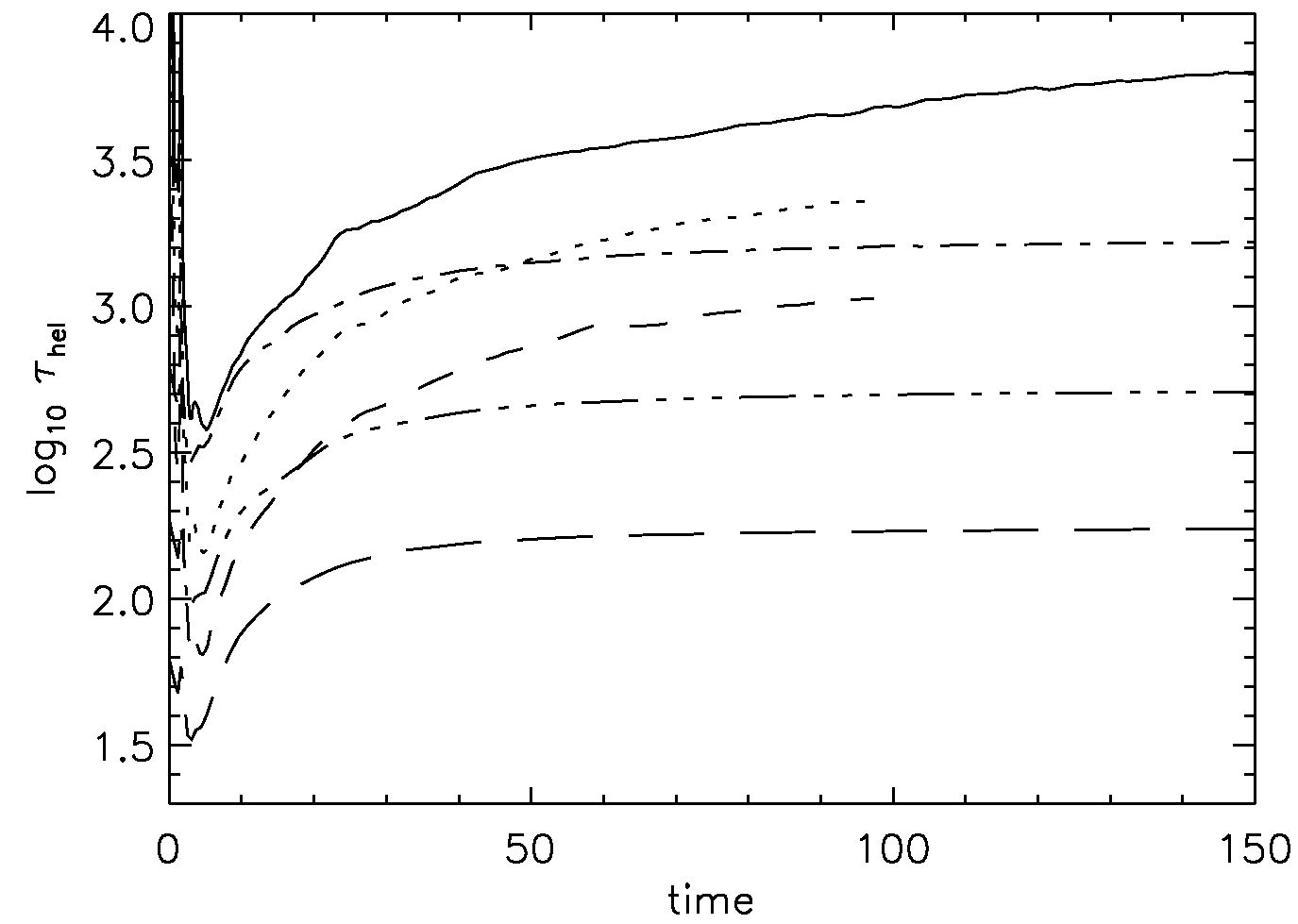}
\caption{{\it Upper panel:} the magnetic energy (thick lines) and helicity (thin lines) against time in six simulations. All have the same initial conditions and are run at resolution $128^3$; the difference is in the diffusivity: three are run with hyperdiffusion and three with standard diffusion, each with magnetic Prandtl numbers of 10, 3 and 1. The viscosity is the same in all simulations. {\it Lower panel:} For the same simulations (with the same line styles), the evolution timescale of the helicity defined as $\tau_{\rm hel}\equiv (\rd \ln |H|/\rd t)^{-1}$. Clearly, the lower the magnetic diffusivity, the better the helicity conservation. The hyperdiffusion scheme also improves helicity conservation.}
\label{fig:helicity}
\end{figure}
\begin{figure}
\includegraphics[width=1.0\hsize,angle=0]{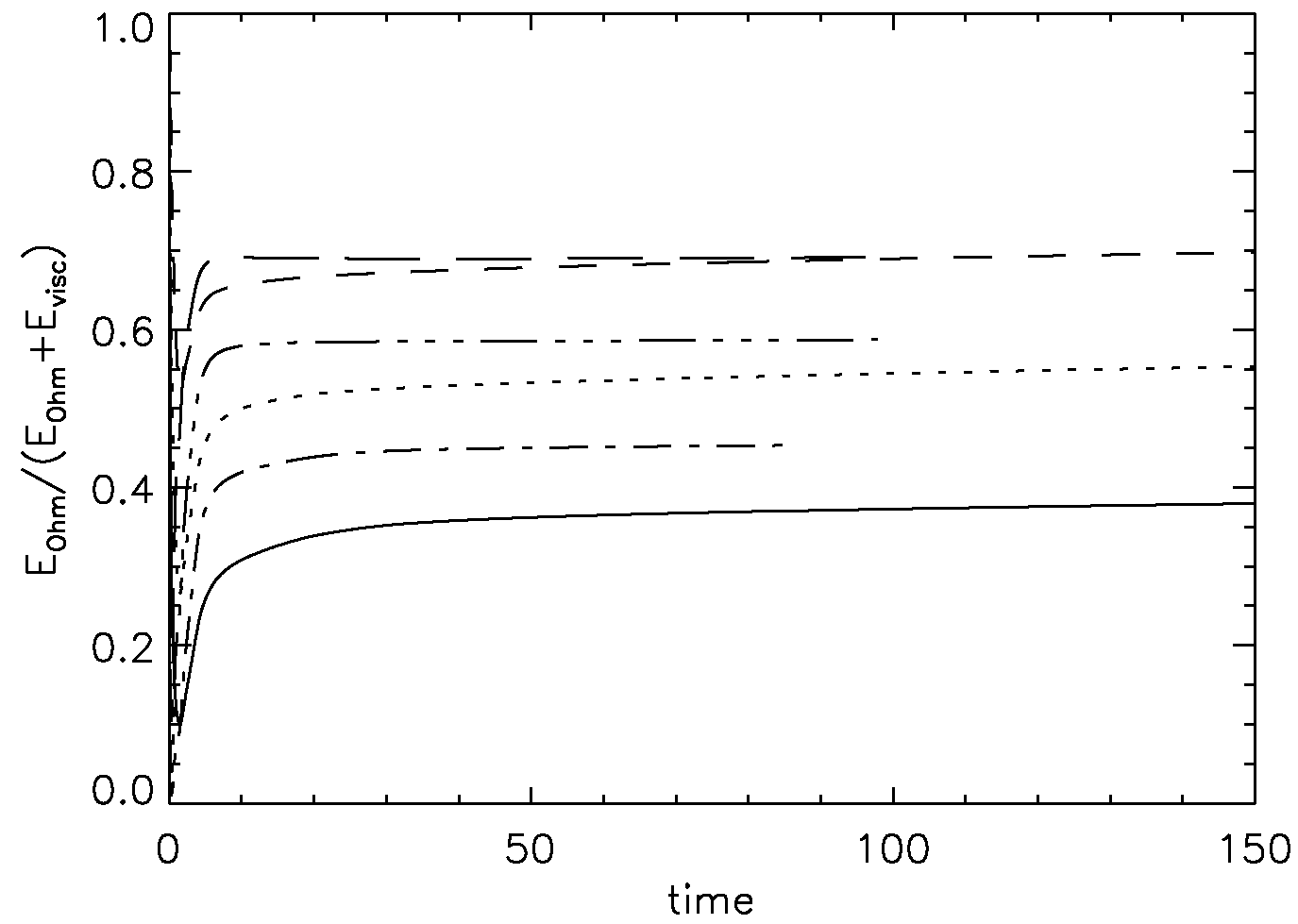}
\caption{For the same six runs as in fig.~\ref{fig:helicity}, the  Ohmic cumulative energy dissipation as a fraction of the sum of Ohmic and viscous cumulative energy dissipations. As expected, the larger one diffusivity with respect to the other, the greater energy it dissipates. There is also some bias towards Ohmic diffusion.}
\label{fig:dissipation}
\end{figure}

\section{Conclusions and discussion}\label{discuss}

Simple simulations were performed in a cubic computational box with an ideal-gas equation of state, uniform initial density and pressure, a smoothly-varying, large-scale initial magnetic field, and no driving of any kind. From these simulations we find the following:
\begin{itemize}
\item Tangential discontinuities form on a dynamical (Alfv\'en) timescale, and topological reconnection then proceeds on a timescale roughly a factor 10 greater than that; gradually the discontinuities become fewer and weaker as an equilibrium is approached. The eventual equilibrium is smooth, containing no discontinuities.
\item Qualititatively, the behaviour is the same in all three plasma-$\beta$ regimes: $\beta\ll1$, $\beta\sim1$ and $\beta\gg1$.
\item {\mk Current sheets form in simulations with fixed and pseudo-vacuum boundaries, as well as periodic.}
\item The addition of an artificial friction force makes no qualitative difference, demonstrating that inertia is irrelevant in the formation of current sheets. {\mk Importantly though, the end-state is quantitatively changed by the friction force.}
\item At higher resolution (or lower diffusivity) the discontinuities are thinner. {\mk Whilst the strongest discontinuities are the same at high and low resolution, at high resolution there is a greater number of weaker discontinuities.}
\item The magnetic helicity $H$ is approximately conserved during the whole process, and at equilibrium, the magnetic energy $E$ is given by  $E=k_{\rm min}H$ where $k_{\rm min}=2\pi/L$ is the minimum possible wavenumber in a computational domain of side $L$.
{\mkb \item In simulations with large helicity, fewer discontinuities form than when helicity is small.}
\item Helicity conservation is better at higher resolution; with normal diffusion (i.e. not hyperdiffusion) the rate of decay of helicity is proportional to magnetic diffusivity.
\end{itemize}

{\mk Tangential discontinuities have already been found in studies of the solar corona, where they are thought to be responsible for much of the heating. The focus of work in this context has been in forming discontinuities by jiggling the field lines around at the boundaries. Parker (1972) argued that there are generally no smooth solutions to such displacements at the boundaries, provided that the displacements are of sufficient magnitude and complexity, a condition it is hard to imagine not being met in reality. All studies of the corona have assumed boundary conditions where the flux is frozen into the boundary and subject to motion parallel to the boundary, mimicking the convective motion we see at the solar photosphere. Usually the field is assumed to be force-free, as the corona has a low plasma-$\beta$.

Tangential discontinuities have also been found in simulations with periodic boundaries and an incompressible equation of state in `reduced MHD', allowing perturbations to a strong uniform guide field \citep{Zhdankin2013}, and also without a guide field \citep{Greco2010}.

 From the simulations presented here, we see that the formation of current sheets is very general: they form in both high- and low-$\beta$ regimes, with various boundaries, with and without a large friction force, and from initial conditions which are topologically either rather close to the eventual equilibrium or very distant. In light of this, one should expect discontinuities to occur in many astrophysical contexts, including in the ISM, where they could explain the observed properties of pulsar scintillations.
 
 In the future one might like to explore more thoroughly the effect of the initial conditions, of which only one type was used here: large scale and smooth. It is conceivable that there are initial conditions which do not produce discontinuities as the magnetic field relaxes, but these might have to be either `special' in some sense, or very close to the eventual equilibrium, a situation it is difficult to imagine arising in reality, except perhaps where an existing equilibrium is `stirred' a just little and then allowed to relax back again. Another possibility is that there is some possible constellation of initial and boundary conditions which produces a equilibrium containing discontinuities.}

Finally, a comment on the observational effects in terms of pulsar scintillations. In the simulations, the gas pressure in the current sheets is generally somewhat different from the surroundings. Of interest for pulsar scintillations is the electron density rather than the pressure, about which these simulations tell us nothing directly, owing to their lack of the relevant thermal physics: they do contain Ohmic heating, but there is no radiative cooling, the thermal conduction is not particularly realistic, and the magnetic Prandtl number is far smaller than in the ISM. The current sheets are heated by reconnection, and will cool on some radiative cooling timescale which should be around $10^5$ yr at a density of $1$ cm$^{-3}$. The longevity of these current sheets is somewhat greater than the dynamical timescale; taking the sheets to be separated by distances of around $0.1$ pc (inferred from observations) and taking the Alfv\'en speed to be 10 km s$^{-1}$, this is also around $10^5$ yr. Given that these timescales are comparable, it is not obvious whether the gas density in current sheets should be lower or higher than in the surroundings. For pulsar scintillations though the sign of the density difference is not important, only the magnitude, which will be a factor of order unity, or somewhat less than unity in locations where the ISM has been left relatively undisturbed for a long time and only weak discontinuities remain.

{\it Acknowledgements.} The author would like to thank Andrei Gruzinov, \AA ke Nordlund, Gene Parker, Ue-Li Pen and Henk Spruit for fruitful discussions and assistance. Some of the simulations were performed on the Sunnyvale machine at CITA, Toronto. Many of the figures (the 3D renderings) were produced with the VAPOR software developed at UCAR in Boulder, Colorado (see www.vapor.ucar.edu).

\bibliography{./Biblio}

\label{lastpage}

\end{document}